\documentclass{aa}  
\usepackage{graphicx}
\usepackage{txfonts}
\usepackage{natbib}
\bibpunct{(}{)}{;}{a}{}{,} 
\newcommand{\siml}{\raise -2.truept\hbox{\rlap{\hbox{$\sim$}}\raise5.truept \hbox{$<$}\ }}
\newcommand{\simg}{\raise -2.truept\hbox{\rlap{\hbox{$\sim$}}\raise5.truept \hbox{$>$}\ }}
\newcommand{\be}{\begin{equation}}
\newcommand{\ee}{\end{equation}}
\newcommand{\ba}{\begin{eqnarray}}
\newcommand{\ea}{\end{eqnarray}}

\newcommand {\hh} {Mpc}

\newcommand {\ks} {km~s$^{-1} \;$}
\newcommand {\kss} {km~s$^{-1}$}

\newcommand {\mqua} {$\times 10^{14}\;M_{\odot} \;$}

\newcommand{\degree}{\ensuremath{\mathrm{^\circ}\;}}

\begin{document}

\title{Velocity segregation effects in galaxy clusters at $0.4 \lesssim z \lesssim 1.5$}

\author{S. Barsanti\inst{\ref{girar}}
\and M. Girardi\inst{\ref{girar},\ref{bivia}}
\and A. Biviano\inst{\ref{bivia}} 
\and S. Borgani\inst{\ref{girar},\ref{bivia},\ref{infn}}
\and M. Annunziatella\inst{\ref{bivia}}
\and M. Nonino\inst{\ref{bivia}}
}

\offprints{S. Barsanti, \email{stefania.jess@gmail.com}}

\institute{
Dipartimento di Fisica, Universit\`a degli Studi di Trieste, via Tiepolo 11, I-34143 Trieste, Italy\label{girar} 
\and 
INAF--Osservatorio Astronomico di Trieste, via G. B. Tiepolo 11, I-34133, Trieste, Italy\label{bivia}
\and
INFN--Sezione di Trieste, via Valerio 2, I-34127 Trieste, Italy\label{infn}
}

\date{Received 27 May 2016/ Accepted 11 August 2016}

\abstract{}{Our study is meant to extend our knowledge of the galaxy
  color and luminosity segregation in velocity space (VCS and VLS, 
  resp.), to clusters at intermediate and high redshift.} {Our sample
  is a collection of 41 clusters in the $0.4 \lesssim z \lesssim 1.5$
  redshift range, for a total of 4172 galaxies, 1674 member galaxies
  within 2$R_{200}$ with photometric or spectroscopic information, as
  taken from the literature. We pay attention to perform homogeneous
  procedures to select cluster members, compute global cluster
  properties, in particular the LOS velocity dispersion $\sigma_{V}$,
  and separate blue from red galaxies.}{We find evidence of VCS in
  clusters out to $z\simeq 0.8$ (at the 97\%-99.99\% c.l., depending on the
  test), in the sense that the blue galaxy population has a 10-20\%
  larger $\sigma_{V}$ than the red galaxy population. Poor or no VCS
  is found in the High-$z$ sample at $z\ge0.8$.  For the first time, we
  detect VLS in non-local clusters and confirm that VLS only affects
  the very luminous galaxies, with brighter galaxies having lower
  velocities. The threshold magnitude of VLS is $\sim m_3+0.5$, where
  $m_3$ is the magnitude of the third brightest cluster galaxy, and
  current data suggest that the threshold value moves to fainter
  magnitudes at higher redshift. We also detect (marginal) evidence of
  VLS for blue galaxies.}  {We conclude that the segregation effects,
  when their study is extended to distant clusters, can be important
  tracers of the galaxy evolution and cluster assembly and discuss the
  poor/no evidence of VCS at high redshift.}

\keywords{
  galaxies: clusters: general -- galaxies: kinematics and dynamics --
  galaxies: evolution -- cosmology: observations}

\titlerunning{Velocity segregation in galaxy clusters}

\maketitle
%

\section{Introduction}
\label{intro}

It is well established that the properties of cluster galaxies differ
from those of field galaxies and that clusters are characterized by
radial gradients. Galaxies in denser, central regions are usually of
earlier morphological type, redder color, and lower star formation
rate. This is the well-known phenomenon of the spatial segregation of
spiral/elliptical galaxies and of blue/red galaxies in local and
distant clusters (e.g., \citealt{Melnick1977}; \citealt{Dressler1980};
\citealt{Whitmore1993}; \citealt{Abraham1996}; \citealt{Dressler1999};
\citealt{Gerken2004}).  More recently, this effect is only questioned
at very high redshift where some authors detect an inversion of the
star formation rate vs. galaxy density (\citealt{Tran2010};
\citealt{Santos2015}, but see \citealt{Ziparo2014}).  The spatial
segregation is a basic observable in the framework of galaxy evolution
and, in particular, of the connection between galaxy evolution and
cluster environment.

A related phenomenon is the segregation of galaxies of different
color/type in velocity space (hereafter VCS). Unfortunately, requiring
a lot of observational effort to measure galaxy redshifts, this effect
is far less known than the spatial segregation.  Since the pioneering
works of \citet{Tammann1972} and \citet{Moss1977}, several studies
have reported significant differences in the velocity
distributions of different galaxy populations. The velocity dispersion
of the population of blue, star-forming galaxies is found to be larger
than that of the population of red, passive galaxies (e.g.,
\citealt{Sodre1989}; \citealt{Biviano1992}; \citealt{Scodeggio1995};
\citealt{Biviano1996}; \citealt{Colless1996}; \citealt{Mohr1996};
\citealt{Biviano1997}; \citealt{Adami1998}; \citealt{Dressler1999};
\citealt{Goto2005}).

\citet{Moss1977} suggested that VCS is an evidence of an infalling
population of field galaxies into the clusters. Large data samples,
obtained stacking galaxies of several clusters, have allowed to trace
the velocity dispersion profiles (VDPs) and obtain new insights on the
issue. \citet{Biviano1997} analyzed the ESO Nearby Abell Cluster
Survey (ENACS -- 107 clusters, see \citealt{Katgert1996}) and inferred
that the kinematical segregation of the emission line galaxies (ELGs)
with respect to the passive galaxy population reflects the time of
infall rather than the virialized condition.  In fact, they found that
the VDP of ELGs is consistent with the fact that ELGs are on more
radial orbits than passive galaxies.  \citet{Carlberg1997} analyzed
the Canadian Network for Observational Cosmology cluster sample (CNOC
-- 16 clusters at medium redshift $z \sim 0.3$, see
\citealt{Yee1996}), finding that the blue galaxy population is
characterized by a larger value of the global velocity dispersion than
the red galaxy population and a different VDP. The difference in the
VDP is an expected consequence of the fact that both populations trace
the same cluster potential with different spatial density profiles.
\citet{Biviano2004} have shown that early and late spectroscopic type
galaxies of ENACS clusters are in equilibrium in the cluster potential
and that late-type galaxies have more radially-elongated orbits.  Some
interesting papers have started to analyze cluster numerical
simulations in the attempt to trace galaxies during their infall into
clusters and relate galaxy properties to their position in the
projected phase space or to the kinematical properties of the galaxy
population, that is velocity distribution and velocity dispersion
(e.g., \citealt{Mahajan2011}; \citealt{Hernandez2014};
\citealt{Haines2015}).  In this context, VCS is therefore an important
observational feature related to galaxy evolution during cluster
assembly.

The presence of VCS is questioned in a few past and
recent studies. Analyzing a sample of six clusters,
\citet{Zabludoff1993} found that the early- and late-type galaxies
have no different velocity dispersions.  The analysis of the Cluster
and Infall Region Nearby Survey (CAIRNS -- 8 clusters at $z<0.05$, see
\citealt{Rines2003}) have shown that the kinematics of star-forming
galaxies in the infall region closely matches that of
absorption-dominated galaxies (\citealt{Rines2005}).
\citet{HwangLee2008} investigated the orbital difference between
early-type and late-type galaxies in ten clusters using data extracted
from SDSS and 2dFGRS data, in four of these they have not found any
difference.  \citet{Rines2013} analyzed the Hectospec Cluster Survey
(HeCS -- 58 clusters with $0.1\leq z \leq 0.3$) obtaining that the
determination of velocity dispersion and dynamical mass is insensitive
to the inclusion of bluer members and that the velocity dispersion of
the ensemble cluster of all galaxies is only 0.8\% larger than that of
the red-sequence galaxies.

The above-mentioned discrepancies can be probably understood taking
into account that the analysis of VCS implies several difficulties and
possible sources of confusion. The member selection is particularly
critical since the effect of including typically blue, field galaxies
can bias the velocity dispersion of the blue population towards higher
values.  Another difficulty is that the amount of VCS, detected so
far, is quantitatively small, accounting for $\sim 10$-20\% (30\% at
most) of the value of the velocity dispersion. For a small velocity
segregation, there is a strong spatial segregation and a decreasing
trend of the VDP, in particular in the case of star-forming galaxies
(\citealt{Biviano2004}). The last two effects combine in such way to
hide the VCS effect when computing global values of the velocity
dispersion, in spite of the positive detection in the VDP (e.g.,
\citealt{Girardi2015}). Indeed, most of the existing positive
detections of VCS have been derived analyzing the VDPs of ensemble
clusters obtained stacking together galaxies of many clusters. The
price to be paid for the large gain in statistics when using ensemble
clusters is that one averages away possible individual behaviors, that
might explain the discrepancies above reported.  Recent and ongoing
cluster catalogs, based on hundreds of member galaxies per cluster
(e.g., \citealt{Owers2011}), will allow us to study the VDP and VCS of
individual clusters.  MACS~J1206.2-0847 is the first cluster of the
CLASH-VLT survey (\citealt{Rosati2014}) where VCS has been analyzed
(\citealt{Girardi2015}).

Moreover, to date, little is known about the velocity segregation in
relation to cluster properties.  For instance, the relation between
VCS and cluster dynamical status has been explored in very few studies
(\citealt{Ribeiro2010}; \citealt{Ribeiro2013}).  The member selection
might be particularly critical in very active clusters, and the
scenario is made more complex by the fact that cluster mergers might
also enhance star formation in galaxies (e.g., \citealt{Caldwell1997};
\citealt{Ferrari2005}; \citealt{Owen2005}).  The dependence of VCS
with redshift is poorly investigated, too. The pioneering study of
\citet{Biviano2009}, based on 18 clusters of the ESO Distant Cluster
Survey (EDisCS -- $z\sim 0.4$-0.8, see \citealt{White2005}), indicates
that VCS is not as pronounced as in local ENACS
clusters. \citet{Crawford2014} analyzed five distant clusters
($0.5<z<0.9$) finding that red sequence, blue cloud, and green valley
galaxies have similar velocity distributions.  To probe of VCS in
distant clusters is on interest also in view of the spectroscopic
survey to be provided by the ESA Euclid mission
\citep{Laureijs2011}. Euclid will provide spectroscopic data for
distant clusters at $0.8<z<1.8$, but only for galaxies with H$\alpha$
lines (\citealt{Sartoris2016} and refs. therein).  This raises the
question as to how velocity dispersions measured using star forming
galaxies compare with those usually measured with red
galaxies. Understanding possible biases in the measurements of
velocity dispersions using different galaxy populations has
implication for cosmological applications of the distribution function
of velocity dispersions (e.g., \citealt{Borgani1997}).  

This paper is devoted to the study of VCS in distant clusters ($0.4
\lesssim z \lesssim 1.5$) and is based on the data of 41 clusters
collected in the literature. To obtain further insights into the
physical processes involved in the velocity segregation, we also
analyzed the possible presence of luminosity segregation in velocity
space (VLS) which is reported in the literature as a minor effect with
respect to VCS (\citealt{Chincarini1977}; \citealt{Biviano1992} and
refs. therein).  In particular, \citet{Biviano1992} have found that
only the most luminous galaxies are segregated in velocity, with
brighter galaxies having lower velocities.  This result has been
confirmed in more recent papers (\citealt{Adami1998};
\citealt{Goto2005}; \citealt{Ribeiro2013}) and in poor group
environments (\citealt{Girardi2003}; \citealt{Ribeiro2010}).  The
observed phenomenology has been explained by physical processes that
transfers kinetic energy from more massive galaxies to less massive
ones. In particular, the dynamical friction (\citealt{Sarazin1986}) is
the most probable mechanism (\citealt{Biviano1992};
\citealt{Mahajan2011}).

The paper is organized as follows. We present our cluster catalog in
Sect.~2. Sections~3 and 4 are devoted to the presentation of member
selection, main cluster properties, and separation between red and
blue galaxy populations. Section~5 concentrates on the analysis of the
VCS and VLS effects, discussed in the following Sect.~6.  We give our
summary and conclusions in Sect.~7.

Unless otherwise stated, we give errors at the 68\% confidence level
(hereafter c.l.). Throughout this paper, we use $H_0=70$ km s$^{-1}$
Mpc$^{-1}$ in a flat cosmology with $\Omega_0=0.3$ and
$\Omega_{\Lambda}=0.7$.

\section{Data sample}
\label{data}

\begin{figure}
\centering
\resizebox{\hsize}{!}{\includegraphics{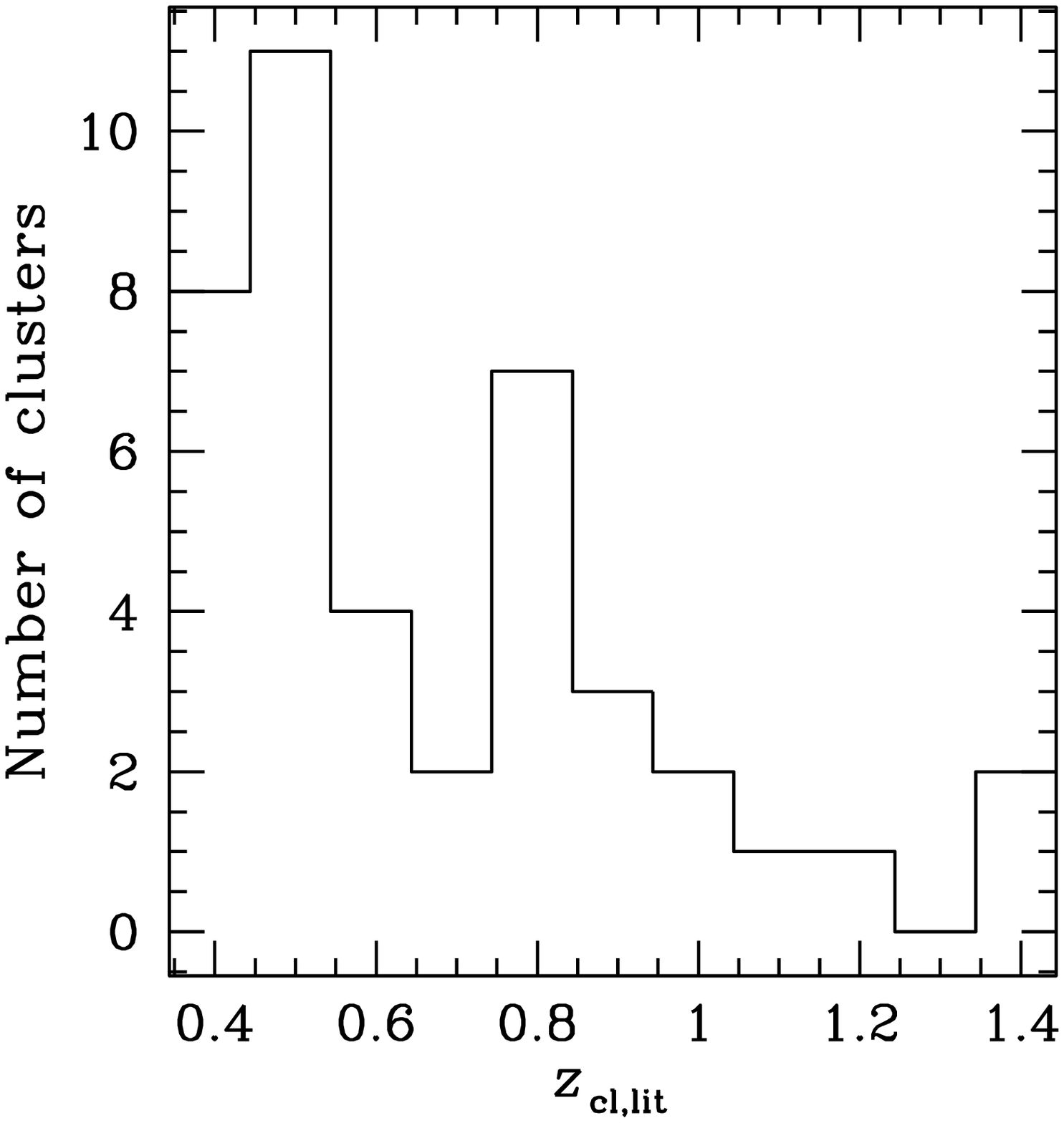}}
\caption
{Distribution of redshifts for the 41 clusters of the sample.}
\label{fighistoz}
\end{figure}

We collected data for clusters with redshift $z \gtrsim 0.4$ and
sampled by at least $20$ galaxies with measured $z$ in the cluster
field.  In order to separate late-type/blue/star-forming galaxies from
early-type/red/passive galaxies, we also required color or spectral
information.  In most cases we used the color information and, in the
following, we refer to the two above classes of galaxies as blue and
red galaxies. In the data collection we also made use of NED
\footnote{The NASA/IPAC Extragalactic Database (NED) is operated by
  the Jet Propulsion Laboratory, California Institute of Technology,
  under contract with the National Aeronautics and Space
  Administration.} looking for cluster data until 2015 June 5.  We
only considered clusters with homogeneous data samples, that is
clusters coming from one author or one collaboration.
Table~\ref{table1} lists the 41 clusters which pass our
requirements  (see the end of Sect.~\ref{redblue}). The cluster
catalog samples the redshift range 0.39-1.46 with a median redshift of
0.58 (see Fig.~\ref{fighistoz}) and is a collection of 4172
galaxies, 100 galaxies per cluster (median value).

\begin{table*}
\caption[]{Cluster sample.}
         \label{table1}
               $$
     \begin{tabular}{l c c r c c}
\hline
\noalign{\smallskip}
\hline
\noalign{\smallskip}
\multicolumn{1}{l}{Cluster name}
&\multicolumn{1}{c}{$z_{\rm cl,lit}$} 
&\multicolumn{1}{c}{Mag} 
&\multicolumn{1}{c}{$N_z$} 
&\multicolumn{1}{c}{$R_{\rm sam}/R_{\rm 200}$} 
&\multicolumn{1}{c}{Refs.}
\\
\hline
\noalign{\smallskip}
\object{MS 0015.9+1609}         & 0.5481 & $g,r$        &  111& 0.92 &6\\
\object{CL 0024+1652}           & 0.3928 & $g,r$        &  130& 0.66 &4\\
\object{RX J0152.7$-$1357}      & 0.8370 & $r,K_{s}$    &  219& 0.57 &3\\
\object{MS 0302.5+1717}         & 0.4249 & $R,I$        &   43& 0.36 &7\\
\object{MS 0302.7+1658}         & 0.4245 & $g,r$        &   96& 0.79 &5\\
\object{CL 0303+1706}           & 0.4184 & $g,r$        &   84& 0.85 &4\\
\object{MS 0451.6$-$0305}       & 0.5398 & $g',r',i',z'$&   70& 0.40 &12\\ 
\object{RDCS J0910+5422}        & 1.1005 & $V,R,i,z,K$  &  156& 2.06 &18\\
\object{CL 0939+4713}           & 0.4060 & $g,r$        &  132& 0.53 &4\\ 
\object{CL J1018.8$-$1211}      & 0.4734 & $B,V,I$      &   71& 0.92 &15,21\\
\object{CL J1037.9$-$1243a}     & 0.4252 & $V,R,I$      &  131& 0.97 &15,21\\
\object{CL J1037.9$-$1243}      & 0.5783 & $V,R,I$      &  131& 1.08 &15,21\\
\object{CL J1040.7$-$1155}      & 0.7043 & $V,R,I$      &  119& 1.36 &10,21\\
\object{CL J1054.4$-$1146}      & 0.6972 & $V,R,I$      &  108& 1.13 &10,21\\
\object{CL J1054.7$-$1245}      & 0.7498 & $V,R,I$      &  100& 1.46 &10,21\\
\object{MS 1054.4$-$0321}       & 0.8307 & $V,i$        &  145& 0.61 &19\\
\object{CL J1059.2$-$1253}      & 0.4564 & $B,V,I$      &   85& 1.05 &15,21\\
\object{CL J1103.7$-$1245a}     & 0.6261 & $V,R,I$      &  178& 2.25 &15,21\\
\object{CL J1103.7$-$1245}      & 0.9580 & $V,R,I$      &  178& 2.30 &20,21\\
\object{RX J1117.4+0743}        & 0.4850 & $g',r'$      &   75& 0.32 &1\\
\object{CL J1138.2$-$1133a}     & 0.4548 & $V,R,I$      &  112& 1.42 &15,21\\
\object{CL J1138.2$-$1133}      & 0.4796 & $V,R,I$      &  112& 0.91 &15,21\\
\object{CL J1216.8$-$1201}      & 0.7943 & $V,R,I$      &  118& 0.78 &10,21\\
\object{RX J1226.9+3332}        & 0.8908 & $r',i',z'$   &  119& 0.62 &12\\
\object{XMMU J1230.3+1339}      & 0.9745 & $r,i,z$      &   23& 4.54 &8,13\\
\object{CL J1232.5$-$1250}      & 0.5414 & $B,V,I$      &   94& 0.51 &10,21\\
\object{RDCS J1252$-$2927}      & 1.2370 & $R,K_{s}$     &  227& 1.18 &2\\
\object{CL J1301.7$-$1139a}     & 0.3969 & $B,V,I$      &   87& 1.47 &15,21\\  
\object{CL J1301.7$-$1139}      & 0.4828 & $B,V,I$      &   87& 0.73 &15,21\\
\object{CL J1324+3011}          & 0.7560 & $B,V,R,I$    &  181& 0.75 &14\\
\object{CL J1353.0$-$1137}      & 0.5882 & $B,V,I$      &   68& 1.05 &15,21\\
\object{CL J1354.2$-$1230}      & 0.7620 & $V,R,I$      &  126& 1.41 &15,21\\ 
\object{CL J1411.1$-$1148}      & 0.5195 & $B,V,I$      &   78& 0.71 &15,21\\
\object{3C 295}                 & 0.4593 & $g,r$        &   35& 0.16 &4\\ 
\object{CL 1601+4253}           & 0.5388 & $g,r$        &   98& 0.78 &4\\
\object{CL J1604+4304}          & 0.8970 & $B,V,R,I$    &   96& 0.89 &16\\
\object{CL J1604+4321}          & 0.9240 & $B,V,R,I$    &  135& 1.06 &16\\	
\object{MS 1621.5+2640}         & 0.4275 & $g,r$        &  262& 2.04 &5\\
\object{RX J1716.6+6708}        & 0.8090 & $r,i,z$      &   37& 0.61 &9\\
\object{XMMXCS J2215.9$-$1738}  & 1.4570 & $I,K_{s}$     &   44& 3.30 &11\\
\object{XMMU J2235.3$-$2557}    & 1.3900 & $J,K_{s}$     &  179& 1.74 &17\\   
\noalign{\smallskip}
\hline
\noalign{\smallskip}
\end{tabular}
$$ \tablebib{ (1)~\citet{Carrasco2007}; (2) \citet{Demarco2007}; (3)
       \citet{Demarco2010}; (4) \citet{Dressler1999}; (5)
       \citet{Ellingson1997}; (6) \citet{Ellingson1998}; (7)
       \citet{Fabricant1994}; (8) \citet{Fassbender2011}; (9)
       \citet{Gioia1999}; (10) \citet{Halliday2004}; (11)
       \citet{Hilton2010}; (12) \citet{Jorgensen2013}; (13)
       \citet{Lerchster2011}; (14) \citet{Lubin2002}; (15)
       \citet{Mil2008}; (16) \citet{Postman2001}; (17)
       \citet{Rosati2009} and data provided by P. Rosati; (18)
       \citet{Tanaka2008}; (19) \citet{Tran2007}; (20)
       \citet{Vulcani2012}; (21) \citet{White2005}.  }
\end{table*}

For each cluster, Table~\ref{table1} lists the name of the cluster
(Col.~1); the redshift as listed in the literature (Col.~2); the
available magnitude information (Col.~3); the number of galaxies with
measured redshift in the field, $N_z$ (Col.~4); the sampling radius
$R_{\rm sam}$, in units of $R_{200}$ (Col.~5); the redshift and
magnitude references (Col.~6).  The sampling radius, which is based on
the estimates of the cluster center and the $R_{200}$ radius
\footnote{The radius $R_{\delta}$ is the radius of a sphere with mass
  overdensity $\delta$ times the critical density at the redshift of
  the galaxy system. Correspondingly, $M_{200}$ is the total mass
  contained within this radius.}  computed in Sect.~\ref{prop}, is listed here
to show the radial extension of the original cluster data sample.

  Our cluster sample results from a collection of data from several
  sources which results in different sampling criteria related to the
  strategy used by the different observers to select spectroscopic
  targets. As a result, the data samples we use differ in their
  photometric properties and completeness limits.  This is not
  expected to affect our results, which are based on kinematics and
  relative comparisons, but we suggest to be cautious in drawing other
  conclusions.  For instance, we report the numbers of cluster
  galaxies only to stress the relative statistical weights of the
  different samples.  Since these numbers are dependent on the
  different (and often poorly known) selection functions of the
  original data sources, they are not representative of the intrinsic
  relative fractions of the different cluster populations.

\section{Member selection and global cluster properties}
\label{prop}

To select cluster members, we applied a two-step procedure introduced
by \citet{Fadda1996} and called ``peak+gap'' (P+G) in more recent
studies (\citealt{Biviano2013}; \citealt{Girardi2015}).  The method is a
combination of the 1D adaptive-kernel method DEDICA
(\citealt{Pisani1993}; see also the Appendix of \citealt{Girardi1996})
and the ``shifting gapper'' method, which uses both position and
velocity information (\citealt{Fadda1996}).  The 1D-DEDICA method is a
non parametric, adaptive method of density reconstruction, optimized as
described in \citet{Pisani1993}. It is used to detect the cluster
peak in the redshift distribution and assign the respective galaxies.
For each cluster, we detected a peak at the cluster
redshift reported by the literature, that is with a difference $\Delta
z <0.003$, with high significant c.l. (i.e. $\ge 99\%$, with the
exception of RX\ J1716.6+6708 at the 98\% c.l.). 
In the few cases where
secondary peaks are also detected, we considered as belonging to the
cluster those peaks $\le$ 2500 \ks apart from the main peak and having
at least a $\ge 25 \%$ of overlapping with the main peak. For each cluster, the
preliminary cluster members are used to compute the center as the mean
position in R.A. and Dec. of the galaxies using the biweight estimator
(ROSTAT software, \citealt{Beers1990}).  The ``shifting gapper''
procedure rejects galaxies that are too far in velocity from the main
body of galaxies and within a fixed radial bin that shifts along the
distance from the cluster center.  The procedure is iterated until the
number of cluster members converges to the final value.  Following
\citet{Fadda1996} we used a gap of $1000$ \ks -- in the cluster
rest-frame -- and a bin of 0.6 \hh, or large enough to include 15
galaxies.

For each cluster, we computed its global properties through a
recursive procedure. First, we estimated the mean cluster redshift,
using the biweight estimator, and the robust estimate of the LOS
velocity dispersion. For robust estimate we mean that we used the
biweight estimator and the gapper estimator for samples with $\ge$ or
$<15$ member galaxies, respectively, following the suggestions of
\citet{Beers1990} and \citet{Girardi1993}. In the computation of
velocity dispersions we also applied the cosmological correction and
the standard correction for velocity errors (\citealt{Danese1980}).
To obtain a first estimate of the radius $R_{200}$ and the cluster mass
$M_{200}$ there contained, we used the theoretical relation between
mass and velocity dispersion of Munari et al. (\citeyear{Munari2013};
Eq.~1), which those authors verified on simulated clusters.  We
considered the galaxies within this first estimate of $R_{200}$ to
recompute the galaxy properties and in particular the final estimate
of $R_{200}$ and $M_{200}$. Using galaxies within this fiducial
estimate of $R_{200}$, we estimated the final cluster properties, that
are the cluster center, the mean redshift $z_{\rm cl}$, and the
velocity dispersion $\sigma_V$, as listed in Table~\ref{table2}. We do
not list the errors on $R_{200}$ (resp. $M_{200}$) since its relative error
is nominally equal (resp. three times) that on $\sigma_V$ considering the
scaling relation, $R_{200}\propto \sigma_V$ (resp. $M_{200}\propto
\sigma_V^3$).  An additional $\sim 10\%$ uncertainty on $M_{200}$
arises from the intrinsic scatter between $M_{200}$ and $\sigma_V$, as
indicated by numerical simulations (\citealt{Munari2013}).

\begin{table*}
\caption[]{Cluster properties.}
         \label{table2}
                $$
         \begin{tabular}{l r r c c c l r r}
\hline
\noalign{\smallskip}
\hline
\noalign{\smallskip}
\multicolumn{1}{l}{Cluster name}
&\multicolumn{1}{c}{$N_{\rm all}$} 
&\multicolumn{1}{c}{$N_{R200}$}
&\multicolumn{1}{c}{$\alpha$ (J2000)}
&\multicolumn{1}{c}{$\delta$ (J2000)}
&\multicolumn{1}{c}{$z_{\rm cl}$} 
&\multicolumn{1}{c}{$\sigma_V$} 
&\multicolumn{1}{r}{$R_{\rm 200}$}
&\multicolumn{1}{r}{$M_{\rm 200}$}
\\
\multicolumn{1}{l}{}
&\multicolumn{1}{c}{}
&\multicolumn{1}{c}{}
&\multicolumn{1}{c}{(hh:mm:ss)}
&\multicolumn{1}{c}{(\degree:\arcmin:\arcsec)}
&\multicolumn{1}{c}{}
&\multicolumn{1}{c}{(km\ $\rm s^{-1}$)} 
&\multicolumn{1}{r}{(Mpc)} 
&\multicolumn{1}{r}{($10^{14} M_{\odot}$)}
\\ 
\hline
\noalign{\smallskip}
MS 0015.9+1609         &  50 &  35 &00:18:31.95 &  +16:25:19.6 &0.5505$\pm$0.0005 & \phantom{11}916$^{\scriptscriptstyle{+139}}_{\scriptscriptstyle{-102}}$ & 1.39 &  5.58\\   
CL 0024+1652           & 100 &  99 &00:26:33.82 &  +17:10:06.7 &0.3936$\pm$0.0003 & \phantom{11}892$^{\scriptscriptstyle{+ 56}}_{\scriptscriptstyle{- 91}}$ & 1.55 &  6.37\\
RX J0152.7$-$1357      & 125 & 124 &01:52:42.20 &$-$13:57:54.5 &0.8359$\pm$0.0004 &\phantom{1}1335$^{\scriptscriptstyle{+ 65}}_{\scriptscriptstyle{- 63}}$ & 1.78 & 16.42\\ 
MS 0302.5+1717         &  28 &  28 &03:05:17.99 &  +17:28:30.0 &0.4242$\pm$0.0004 & \phantom{11}666$^{\scriptscriptstyle{+ 62}}_{\scriptscriptstyle{- 72}}$ & 1.13 &  2.60\\                   
MS 0302.7+1658         &  34 &  33 &03:05:31.68 &  +17:10:06.1 &0.4248$\pm$0.0005 & \phantom{11}793$^{\scriptscriptstyle{+120}}_{\scriptscriptstyle{- 87}}$ & 1.35 &  4.39\\ 
CL 0303+1706           &  46 &  44 &03:06:14.40 &  +17:18:00.3 &0.4188$\pm$0.0004 & \phantom{11}804$^{\scriptscriptstyle{+110}}_{\scriptscriptstyle{-139}}$ & 1.37 &  4.59\\
MS 0451.6$-$0305       &  44 &  44 &04:54:10.96 &$-$03:01:07.8 &0.5401$\pm$0.0006 &\phantom{1}1242$^{\scriptscriptstyle{+ 72}}_{\scriptscriptstyle{-106}}$ & 1.98 & 15.77\\
RDCS J0910+5422        &  23 &  16 &09:10:45.70 &  +54:22:22.4 &1.1004$\pm$0.0006 & \phantom{11}705$^{\scriptscriptstyle{+153}}_{\scriptscriptstyle{-139}}$ & 0.81 &  2.08\\
CL 0939+4713           &  70 &  70 &09:42:58.68 &  +46:58:59.9 &0.4060$\pm$0.0005 &\phantom{1}1156$^{\scriptscriptstyle{+ 96}}_{\scriptscriptstyle{- 86}}$ & 1.99 & 13.76\\
CL J1018.8$-$1211      &  34 &  27 &10:18:46.71 &$-$12:12:23.1 &0.4734$\pm$0.0004 & \phantom{11}532$^{\scriptscriptstyle{+ 84}}_{\scriptscriptstyle{- 60}}$ & 0.94 &  1.59\\
CL J1037.9$-$1243a     &  47 &  38 &10:37:49.28 &$-$12:45:12.0 &0.4251$\pm$0.0003 & \phantom{11}521$^{\scriptscriptstyle{+ 59}}_{\scriptscriptstyle{- 32}}$ & 0.86 &  1.15\\
CL J1037.9$-$1243      &  19 &  17 &10:37:53.15 &$-$12:43:44.1 &0.5785$\pm$0.0005 & \phantom{11}564$^{\scriptscriptstyle{+248}}_{\scriptscriptstyle{-180}}$ & 0.88 &  1.44\\         
CL J1040.7$-$1155      &  30 &  17 &10:40:40.27 &$-$11:56:12.6 &0.7045$\pm$0.0004 & \phantom{11}517$^{\scriptscriptstyle{+ 80}}_{\scriptscriptstyle{- 44}}$ & 0.79 &  1.21\\
CL J1054.4$-$1146      &  49 &  33 &10:54:25.59 &$-$11:46:42.5 &0.6981$\pm$0.0004 & \phantom{11}607$^{\scriptscriptstyle{+120}}_{\scriptscriptstyle{- 83}}$ & 0.85 &  1.53\\                 	                   	  
CL J1054.7$-$1245      &  36 &  23 &10:54:43.48 &$-$12:46:23.8 &0.7504$\pm$0.0004 & \phantom{11}525$^{\scriptscriptstyle{+137}}_{\scriptscriptstyle{- 72}}$ & 0.74 &  1.07\\
MS 1054.4$-$0321       & 143 & 140 &10:57:00.47 &$-$03:37:32.9 &0.8307$\pm$0.0003 &\phantom{1}1113$^{\scriptscriptstyle{+ 78}}_{\scriptscriptstyle{- 57}}$ & 1.49 &  9.54\\
CL J1059.2$-$1253      &  42 &  39 &10:59:08.61 &$-$12:53:51.6 &0.4564$\pm$0.0003 & \phantom{11}523$^{\scriptscriptstyle{+ 60}}_{\scriptscriptstyle{- 49}}$ & 0.87 &  1.24\\    
CL J1103.7$-$1245a     &  17 &  10 &11:03:35.99 &$-$12:46:45.0 &0.6261$\pm$0.0004 & \phantom{11}357$^{\scriptscriptstyle{+ 48}}_{\scriptscriptstyle{-189}}$ & 0.60 &  0.50\\ 
CL J1103.7$-$1245      &  22 &   9 &11:03:44.69 &$-$12:45:34.1 &0.9580$\pm$0.0006 & \phantom{11}448$^{\scriptscriptstyle{+155}}_{\scriptscriptstyle{-116}}$ & 0.49 &  0.39\\
RX J1117.4+0743        &  37 &  37 &11:17:26.24 &  +07:43:50.4 &0.4857$\pm$0.0008 &\phantom{1}1426$^{\scriptscriptstyle{+219}}_{\scriptscriptstyle{- 97}}$ & 2.34 & 24.65\\
CL J1138.2$-$1133a     &  14 &  14 &11:38:06.09 &$-$11:36:15.1 &0.4546$\pm$0.0005 & \phantom{11}510$^{\scriptscriptstyle{+ 74}}_{\scriptscriptstyle{- 63}}$ & 0.85 &  1.15\\
CL J1138.2$-$1133      &  49 &  48 &11:38:09.86 &$-$11:33:35.3 &0.4797$\pm$0.0003 & \phantom{11}712$^{\scriptscriptstyle{+ 65}}_{\scriptscriptstyle{- 86}}$ & 1.17 &  3.08\\
CL J1216.8$-$1201      &  66 &  65 &12:16:44.59 &$-$12:01:20.3 &0.7939$\pm$0.0004 &\phantom{1}1004$^{\scriptscriptstyle{+ 79}}_{\scriptscriptstyle{- 68}}$ & 1.37 &  7.16\\
RX J1226.9+3332        &  50 &  46 &12:26:58.34 &  +33:32:52.6 &0.8912$\pm$0.0005 &\phantom{1}1039$^{\scriptscriptstyle{+116}}_{\scriptscriptstyle{-107}}$ & 1.34 &  7.49\\
XMMU J1230.3+1339      &  13 &   8 &12:30:17.93 &  +13:39:03.8 &0.9755$\pm$0.0007 & \phantom{11}548$^{\scriptscriptstyle{+206}}_{\scriptscriptstyle{- 98}}$ & 0.75 &  1.44\\
CL J1232.5$-$1250      &  54 &  54 &12:32:30.76 &$-$12:50:41.1 &0.5418$\pm$0.0005 &\phantom{1}1089$^{\scriptscriptstyle{+108}}_{\scriptscriptstyle{-100}}$ & 1.73 & 10.62\\
RDCS J1252$-$2927      &  38 &  28 &12:52:54.40 &$-$29:27:17.4 &1.2367$\pm$0.0005 & \phantom{11}789$^{\scriptscriptstyle{+ 96}}_{\scriptscriptstyle{- 89}}$ & 0.85 &  2.85\\
CL J1301.7$-$1139a     &  17 &  17 &13:01:36.74 &$-$11:39:24.9 &0.3969$\pm$0.0003 & \phantom{11}388$^{\scriptscriptstyle{+ 74}}_{\scriptscriptstyle{- 64}}$ & 0.67 &  0.52\\ 
CL J1301.7$-$1139      &  37 &  31 &13:01:37.08 &$-$11:39:33.0 &0.4831$\pm$0.0004 & \phantom{11}704$^{\scriptscriptstyle{+ 90}}_{\scriptscriptstyle{- 83}}$ & 1.14 &  2.83\\
CL J1324+3011          &  44 &  42 &13:24:48.68 &  +30:11:20.2 &0.7548$\pm$0.0005 & \phantom{11}871$^{\scriptscriptstyle{+139}}_{\scriptscriptstyle{- 97}}$ & 1.20 &  4.54\\            
CL J1353.0$-$1137      &  21 &  17 &13:53:02.15 &$-$11:37:10.9 &0.5880$\pm$0.0005 & \phantom{11}615$^{\scriptscriptstyle{+257}}_{\scriptscriptstyle{-127}}$ & 0.92 &  1.69\\
CL J1354.2$-$1230      &  23 &  14 &13:54:10.16 &$-$12:31:03.1 &0.7593$\pm$0.0005 & \phantom{11}489$^{\scriptscriptstyle{+ 88}}_{\scriptscriptstyle{- 45}}$ & 0.74 &  1.07\\         
CL J1411.1$-$1148      &  25 &  24 &14:11:04.30 &$-$11:48:10.1 &0.5196$\pm$0.0005 & \phantom{11}784$^{\scriptscriptstyle{+145}}_{\scriptscriptstyle{-103}}$ & 1.26 &  4.02\\
3C 295                 &  25 &  25 &14:11:20.06 &  +52:12:16.5 &0.4593$\pm$0.0011 &\phantom{1}1677$^{\scriptscriptstyle{+192}}_{\scriptscriptstyle{-147}}$ & 2.80 & 40.72\\
CL 1601+4253           &  55 &  53 &16:03:09.84 &  +42:45:13.1 &0.5401$\pm$0.0003 & \phantom{11}697$^{\scriptscriptstyle{+ 82}}_{\scriptscriptstyle{- 84}}$ & 1.11 &  2.79\\
CL J1604+4304          &  16 &  13 &16:04:24.70 &  +43:04:48.2 &0.8983$\pm$0.0007 & \phantom{11}683$^{\scriptscriptstyle{+282}}_{\scriptscriptstyle{-139}}$ & 0.88 &  2.12\\             		  
CL J1604+4321          &  37 &  35 &16:04:34.41 &  +43:21:01.6 &0.9220$\pm$0.0004 & \phantom{11}669$^{\scriptscriptstyle{+231}}_{\scriptscriptstyle{-119}}$ & 0.91 &  2.42\\        	                    	  
MS 1621.5+2640         & 104 &  54 &16:23:37.03 &  +26:35:08.7 &0.4257$\pm$0.0003 & \phantom{11}757$^{\scriptscriptstyle{+ 84}}_{\scriptscriptstyle{- 75}}$ & 1.29 &  3.81\\                    	                    	                   
RX J1716.6+6708        &  31 &  28 &17:16:48.86 &  +67:08:22.3 &0.8063$\pm$0.0008 &\phantom{1}1299$^{\scriptscriptstyle{+136}}_{\scriptscriptstyle{-158}}$ & 1.76 & 15.39\\                                                                 	  
XMMXCS J2215.9$-$1738  &  41 &  27 &22:15:58.75 &$-$17:37:57.9 &1.4569$\pm$0.0005 & \phantom{11}745$^{\scriptscriptstyle{+120}}_{\scriptscriptstyle{- 86}}$ & 0.70 &  2.01\\             		                               	  
XMMU J2235.3$-$2557    &  30 &  20 &22:35:20.81 &$-$25:57:22.0 &1.3905$\pm$0.0007 & \phantom{11}910$^{\scriptscriptstyle{+187}}_{\scriptscriptstyle{- 82}}$ & 0.96 &  4.86\\   
\noalign{\smallskip}                  
\hline                               
\noalign{\smallskip}
\end{tabular}
$$ \tablefoot{Column 1: cluster name; Col.~2: the number of all
           fiducial cluster members, $N_{\rm all}$; Col.~3: the
           number of member galaxies contained within $R_{200}$, $N_{R200}$; Cols.~4 and 5: the cluster center; Col.~6: the
           mean redshift, $z_{\rm cl}$; Col.~7: the LOS velocity
           dispersion, $\sigma_V$; Cols.~8 and 9: $R_{200}$ and
           $M_{200}$.  
}
\end{table*}

Note that, 23 clusters are not sampled out to $R_{200}$ (see
Table~\ref{table1}). We used the other 18 well sampled clusters to
verify that this undersampling does not introduce any bias in our
estimate of the velocity dispersion and, consequently, of $R_{200}$
and $M_{200}$. For these 18 clusters, we compared the distribution of
the velocity dispersions computed within 0.5$R_{200}$ and that within
$R_{200}$. We obtained no significant evidence of difference according
to the Kolmogorov-Smirnov test (hereafter KS--test; see, e.g.,
\citealt{Lederman1984}), and according to two more sensitive tests,
the Sign and Wilcoxon Signed--ranks tests (hereafter S-- and W--tests,
e.g., \citealt{Siegel1956}).

\section{Populations of red and blue galaxies}
\label{redblue}

\begin{figure}
\centering
\resizebox{\hsize}{!}{\includegraphics{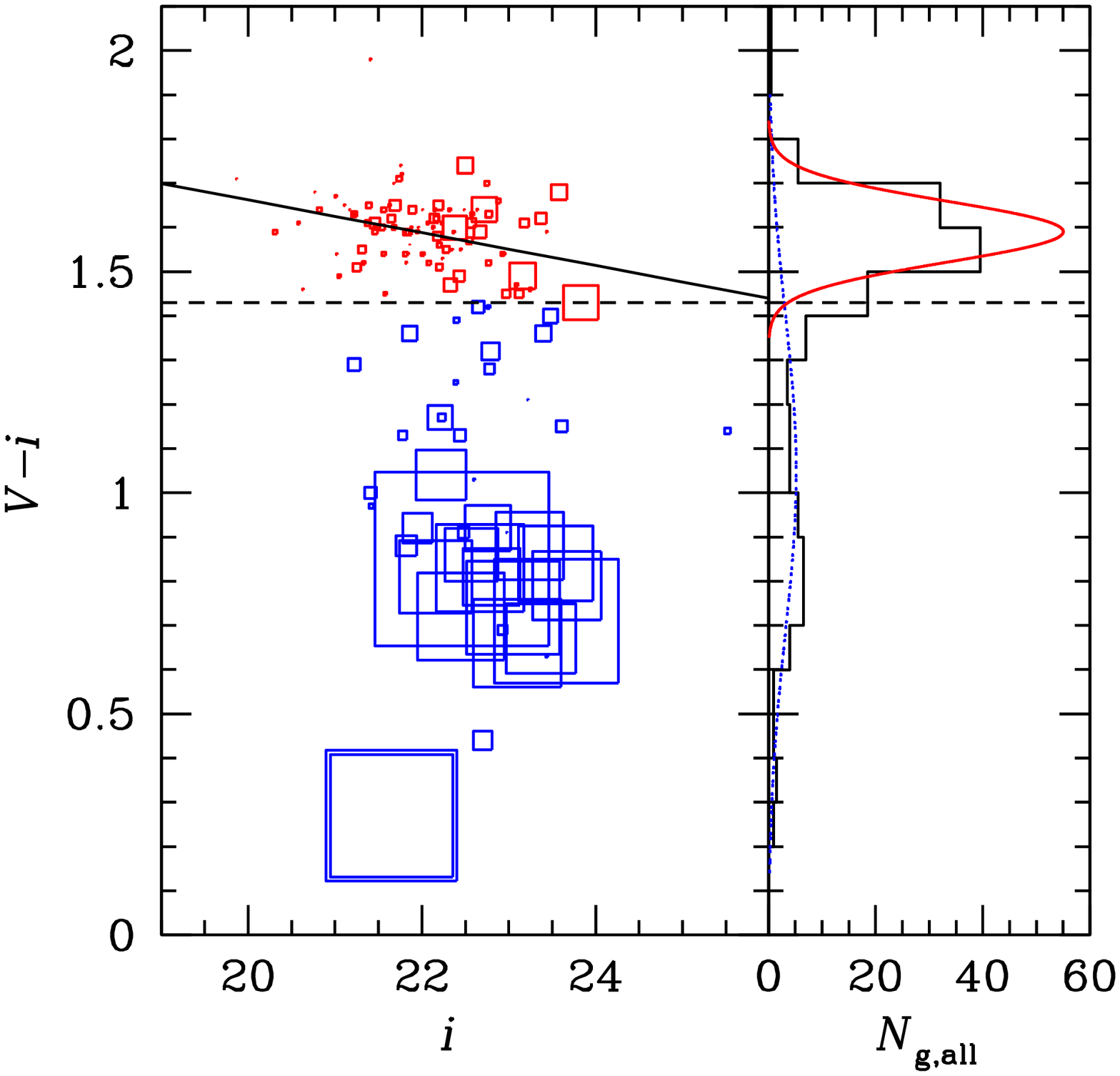}}
\caption
{Separation between red and blue galaxies in the
  MS~1054.4$-$0321 cluster.  {\em Left panel}: color-magnitude
  diagram. The horizontal dashed line indicates the color cut as
  obtained using the $V-i$ colors for member galaxies.
For this well sampled cluster we also show
  the red sequence line as fitted with a two sigma procedure applied
  to the red galaxies (black solid line).
Larger sizes of the symbols for larger EW[OII] show the good
  agreement between the photometric and spectroscopic methods to
  separate red/passive from blue/star-forming galaxy populations.
{\em Right panel}: distribution of
  $V-i$ colors for member galaxies.  The two Gaussians are obtained
  through the KMM method and allow us to define the color cut in 
the left panel.}
\label{figcol}
\end{figure}

In order to separate red/passive from blue/star-forming galaxies, we
used a color based procedure. As much as possible, we considered the
two magnitude bands in such way that the Balmer break at the cluster
redshift lies roughly between the two filters (see Fig.~18 of
\citealt{Westra2010}).  The color distribution is analyzed using the
Kaye's mixture model (KMM) method, as implemented by
\citet{Ashman1994}, to detect the color bimodality and define the
respective group partition and, consequently, the value of the color
cut (see Fig.~\ref{figcol} as an example).

The KMM procedure fails in detecting a significant bimodality in eight
clusters, typically those having very few available data, and we used
alternative procedures.  In four clusters, to define the color cut, we
adopted the intermediate value between the typical color of red and
blue galaxies at $z=0$ (\citealt{Fukugita1995}), conveniently shifted
at the cluster redshift using the k- and evolutionary corrections by
\citet{Poggianti1997}. In the four distant clusters, for which the
magnitude corrections are less reliable, we preferred to use
spectroscopic features to separate red/passive from blue/star-forming
galaxies. We defined as star-forming galaxies those with the [OII]
emission line in their spectrum. When EW[OII] measures were available,
we defined as star-forming galaxies those with EW[OII]$\ge 15$
\AA\ (see, e.g., \citealt{Postman1998}; \citealt{Hammer1997}). For the
eleven clusters where both good magnitude and spectroscopic
information are available, the location of ELGs in the color-mag
diagram supports the good agreement of the photometric- and
spectroscopic-based methods (see Fig.~\ref{figcol} -- left panel --
for an example).  For each cluster, Table~\ref{table3} lists: the
number of member galaxies with measured magnitudes, $N_m$ (Col.~2);
the color and magnitude used in our analysis (Cols.~3 and 4); the
adopted color cut (Col.~5). In the case of the spectroscopic based
separation, $N_m$ refers to the number of galaxies with available
EW[OII] information.  The relevant information about the reference
sources of magnitudes is listed in Table~\ref{table1}.

\begin{table}
\caption[]{Color cuts to separate blue from red galaxies.}
         \label{table3}
                $$
         \begin{tabular}{l r c c c}
\hline
\noalign{\smallskip}
\hline
\noalign{\smallskip}
\multicolumn{1}{l}{Cluster name}
&\multicolumn{1}{c}{$N_m$}
&\multicolumn{1}{c}{Color}
&\multicolumn{1}{c}{Mag}
&\multicolumn{1}{c}{Cut}
\\
\hline
\noalign{\smallskip}
MS 0015.9+1609         	   & 50 & $g-r$ & $r$ & 1.20\\
CL 0024+1652               & 78 & $g-r$ & $r$ & 1.20\\
RX J0152.7$-$1357      	   &106 & $r-K_{\rm s}$ & $K_{\rm s}$ & 2.52\\ 
MS 0302.5+1717         	   & 28 & $R-I$ & $I$ & 0.76\\ 	      
MS 0302.7+1658$^{a}$  	   & 34 & $g-r$ & $r$ & 1.26\\
CL 0303+1706           	   & 37 & $g-r$ & $r$ & 1.37\\
MS 0451.6$-$0305       	   & 44 & $g'-r'$ & $r'$ & 1.76\\ 
RDCS J0910+5422        	   & 23 & $i-z$ & $z$ & 0.82\\
CL 0939+4713           	   & 48 & $g-r$ & $r$ & 1.13\\
CL J1018.8$-$1211          & 34 & $B-I$ & $I$ & 2.94\\
CL J1037.9$-$1243a         & 47 & $V-I$ & $I$ & 1.60\\ 
CL J1037.9$-$1243          & 19 & $V-R$ & $R$ & 1.08\\
CL J1040.7$-$1155          & 30 & $R-I$ & $I$ & 1.02\\
CL J1054.4$-$1146          & 49 & $R-I$ & $I$ & 2.15\\ 	      
CL J1054.7$-$1245          & 35 & $R-I$ & $I$ & 1.11\\ 
MS 1054.4$-$0321       	   &142 & $V-i$ & $i$ & 1.43\\
CL J1059.2$-$1253          & 41 & $B-I$ & $I$ & 3.20\\
CL J1103.7$-$1245a$^{a}$   & 15 & $V-I$ & $I$ & 2.20\\ 
CL J1103.7$-$1245      	   & 22 & $R-I$ & $I$ & 1.20\\
RX J1117.4+0743$^{a}$  	   & 37 & $g'-r'$ & $r'$ & 1.78\\
CL J1138.2$-$1133a         & 14 & $V-I$ & $I$ & 1.88\\ 
CL J1138.2$-$1133          & 49 & $V-I$ & $I$ & 1.69\\
CL J1216.8$-$1201          & 66 & $R-I$ & $I$ & 1.28\\ 
RX J1226.9+3332        	   & 50 & $i'-z'$ & $z'$ &0.63\\ 
XMMU J1230.3+1339      	   & 13 & $i-z$ & $z$ & 0.37\\
CL J1232.5$-$1250          & 54 & $B-I$ & $I$ & 3.78\\
RDCS J1252$-$2927$^{b}$    & 38 & $-$   & $-$ &$-$\\
CL J1301.7$-$1139a     	   & 17 & $B-I$ & $I$ & 3.02\\
CL J1301.7$-$1139      	   & 37 & $B-I$ & $I$ & 3.20\\
CL J1324+3011          	   & 44 & $R-I$ & $I$ & 1.21\\
CL J1353.0$-$1137      	   & 21 & $V-I$ & $I$ & 2.16\\
CL J1354.2$-$1230      	   & 22 & $V-I$ & $I$ & 2.11\\
CL J1411.1$-$1148      	   & 25 & $B-I$ & $I$ & 3.58\\
3C 295$^{a}$               & 25 & $g-r$ & $r$ & 1.40\\ 
CL 1601+4253           	   & 50 & $g-r$ & $r$ & 1.20\\
CL J1604+4304          	   & 15 & $B-R$ & $R$ & 1.05\\	      
CL J1604+4321          	   & 37 & $B-R$ & $R$ & 2.00\\	       	       
MS 1621.5+2640         	   &104 & $g-r$ & $r$ & 1.27\\ 	      
RX J1716.6+6708$^{b}$  	   & 31 &  $-$  & $-$ & $-$\\ 		       	       
XMMXCS J2215.9$-$1738$^{b}$& 41 & $-$   & $-$ &$-$\\ 
XMMU J2235.3$-$2557$^{b}$  & 29 & $-$   & $-$ &$-$\\ 
\noalign{\smallskip}                  
\hline                               
\noalign{\smallskip}
\end{tabular}
$$ \tablefoot{$^{(a)}$ Color cut based on the typical color of
           galaxies (\citealt{Fukugita1995}); $^{(b)}$ Selection
           based on the EW[OII] information.  }
\end{table}

\begin{figure}
\centering
\resizebox{\hsize}{!}{\includegraphics{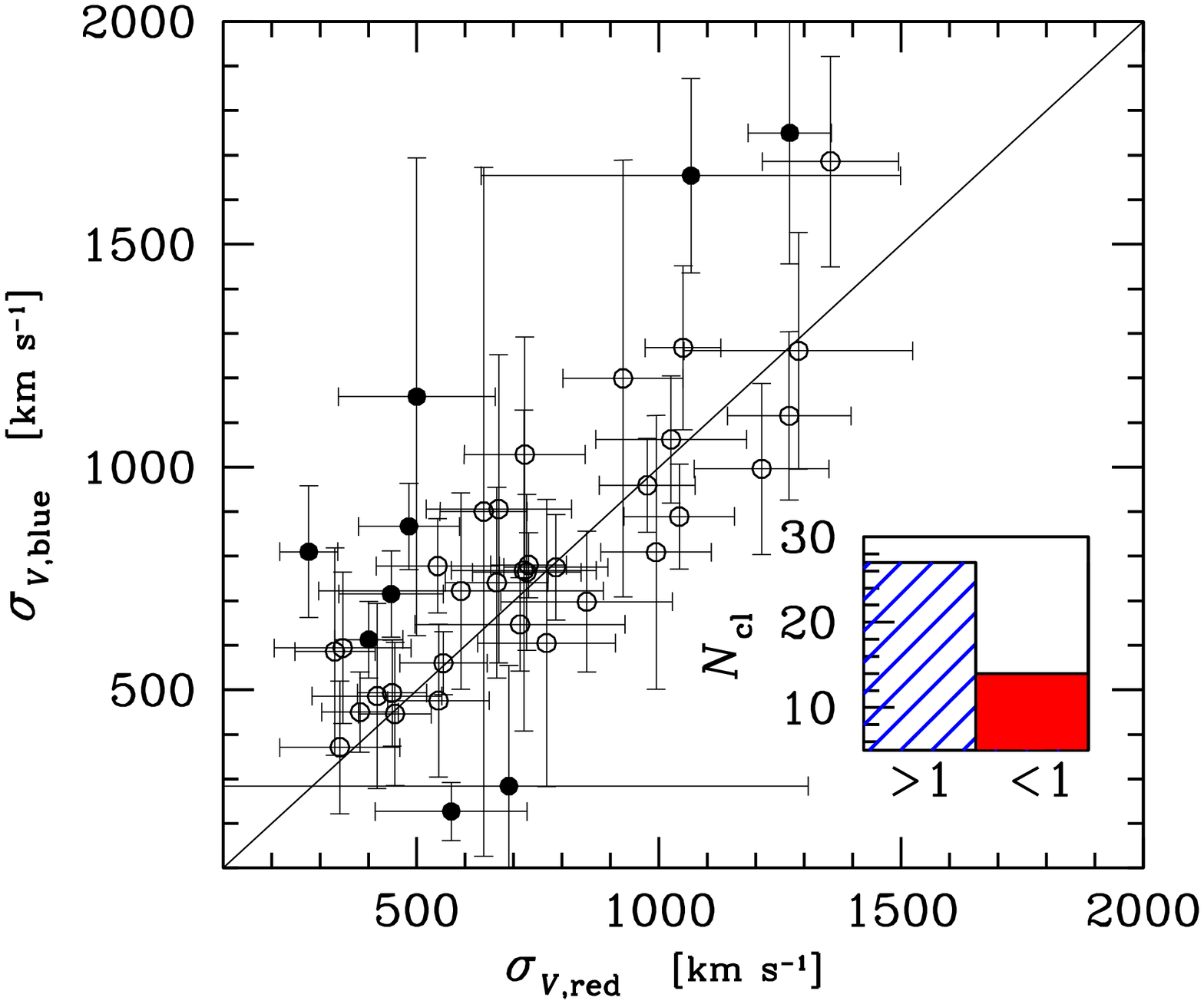}}
\caption
{Velocity dispersion of the red galaxy population vs velocity
  dispersion of the blue galaxy population for all clusters of our
  sample. The filled symbols highlight the clusters where the two
  values are different according to the F--test (at a c.l. $\ge 90\%$,
  see Table~\ref{table5}). Inset shows the number of clusters with
  $\sigma_{V,\rm blue} / \sigma_{V,\rm red} >1$ (blue shaded bar) and the
  number of clusters with $\sigma_{V,\rm blue} / \sigma_{V,\rm red} < 1$
  (red solid bar).}
\label{figss}
\end{figure}

We considered the cluster regions within 2$R_{200}$.  Out to
2$R_{200}$ the cluster density and mass profiles are a reasonable
extrapolation of those determined within $R_{200}$
(\citealt{Biviano2003}).  A requirement of our catalog is that each
cluster is sampled at least with four red and four blue galaxy members
within 2$R_{200}$.  Generally, the clusters in our catalog are much
better sampled (see Cols.~2 and 4 in Table~\ref{table4}), with 20 red
galaxies and 14 blue galaxies per cluster (median
values). Table~\ref{table4} lists the velocity dispersions of the red
and blue galaxies within 2$R_{200}$. The values of $\sigma_{V,\rm
  red}$ and $\sigma_{V,\rm blue}$ correlate at the $>99.99\%$ c.l.
according to the Spearman test (coefficient value of 0.70, see
Fig.~\ref{figss}).

\begin{table}
\caption[]{Velocity dispersions of red and blue populations.}
         \label{table4}
                $$
         \begin{tabular}{l r l r l}
\hline
\noalign{\smallskip}
\hline
\noalign{\smallskip}
\multicolumn{1}{l}{Cluster name}
&\multicolumn{1}{c}{$N_{\rm r}$}
&\multicolumn{1}{c}{$\sigma_{V,\rm red}$}  
&\multicolumn{1}{c}{$N_{\rm b}$} 
&\multicolumn{1}{c}{$\sigma_{V,\rm blue}$}
\\
\multicolumn{1}{l}{}
&\multicolumn{1}{c}{}
&\multicolumn{1}{c}{(km\ $\rm s^{-1}$)}
&\multicolumn{1}{c}{}
&\multicolumn{1}{c}{(km\ $\rm s^{-1}$)}
\\
\hline
\noalign{\smallskip}
MS 0015.9+1609         & 39&  \phantom{11}994$^{\scriptscriptstyle{+138}}_{\scriptscriptstyle{- 87}}$ & 11&  \phantom{11}809$^{\scriptscriptstyle{+538}}_{\scriptscriptstyle{-140}}$\\ 
CL 0024+1652           & 49& \phantom{1}1042$^{\scriptscriptstyle{+115}}_{\scriptscriptstyle{-110}}$ & 29&  \phantom{11}889$^{\scriptscriptstyle{+108}}_{\scriptscriptstyle{-116}}$\\   
RX J0152.7$-$1357      & 91& \phantom{1}1270$^{\scriptscriptstyle{+ 85}}_{\scriptscriptstyle{- 68}}$ & 15& \phantom{1}1750$^{\scriptscriptstyle{+396}}_{\scriptscriptstyle{-289}}$\\
MS 0302.5+1717         & 24&  \phantom{11}638$^{\scriptscriptstyle{+ 63}}_{\scriptscriptstyle{- 82}}$ &  4&  \phantom{11}900$^{\scriptscriptstyle{+482}}_{\scriptscriptstyle{-420}}$\\                    	  
MS 0302.7+1658         & 24&  \phantom{11}669$^{\scriptscriptstyle{+166}}_{\scriptscriptstyle{-113}}$ & 10&  \phantom{11}906$^{\scriptscriptstyle{+623}}_{\scriptscriptstyle{-282}}$\\ 
CL 0303+1706           & 21&  \phantom{11}690$^{\scriptscriptstyle{+309}}_{\scriptscriptstyle{-320}}$ & 16&  \phantom{11}284$^{\scriptscriptstyle{+132}}_{\scriptscriptstyle{-150}}$\\ 
MS 0451.6$-$0305       & 28& \phantom{1}1269$^{\scriptscriptstyle{+124}}_{\scriptscriptstyle{-102}}$ & 16& \phantom{1}1115$^{\scriptscriptstyle{+161}}_{\scriptscriptstyle{-169}}$\\
RDCS J0910+5422        & 16&  \phantom{11}721$^{\scriptscriptstyle{+153}}_{\scriptscriptstyle{-131}}$ &  6&  \phantom{11}768$^{\scriptscriptstyle{+267}}_{\scriptscriptstyle{-252}}$\\ 
CL 0939+4713           & 34& \phantom{1}1212$^{\scriptscriptstyle{+147}}_{\scriptscriptstyle{-117}}$ & 14&  \phantom{11}996$^{\scriptscriptstyle{+207}}_{\scriptscriptstyle{-156}}$\\
CL J1018.8$-$1211      & 19&  \phantom{11}449$^{\scriptscriptstyle{+ 85}}_{\scriptscriptstyle{- 54}}$ & 15&  \phantom{11}493$^{\scriptscriptstyle{+180}}_{\scriptscriptstyle{- 85}}$\\
CL J1037.9$-$1243a     & 26&  \phantom{11}555$^{\scriptscriptstyle{+133}}_{\scriptscriptstyle{- 63}}$ & 21&  \phantom{11}560$^{\scriptscriptstyle{+ 76}}_{\scriptscriptstyle{- 54}}$\\ 
CL J1037.9$-$1243      &  6&  \phantom{11}347$^{\scriptscriptstyle{+113}}_{\scriptscriptstyle{- 72}}$ & 13&  \phantom{11}594$^{\scriptscriptstyle{+231}}_{\scriptscriptstyle{-274}}$\\
CL J1040.7$-$1155      & 12&  \phantom{11}383$^{\scriptscriptstyle{+ 89}}_{\scriptscriptstyle{- 45}}$ & 18&  \phantom{11}450$^{\scriptscriptstyle{+ 82}}_{\scriptscriptstyle{- 59}}$\\
CL J1054.4$-$1146      & 28&  \phantom{11}447$^{\scriptscriptstyle{+114}}_{\scriptscriptstyle{- 75}}$ & 21& \phantom{11}715$^{\scriptscriptstyle{+112}}_{\scriptscriptstyle{- 67}}$\\                 	  
CL J1054.7$-$1245      & 23&  \phantom{11}545$^{\scriptscriptstyle{+130}}_{\scriptscriptstyle{- 76}}$ & 11&  \phantom{11}476$^{\scriptscriptstyle{+286}}_{\scriptscriptstyle{-105}}$\\ 
MS 1054.4$-$0321       & 96& \phantom{1}1050$^{\scriptscriptstyle{+ 76}}_{\scriptscriptstyle{- 74}}$ & 46& \phantom{1}1268$^{\scriptscriptstyle{+203}}_{\scriptscriptstyle{-148}}$\\ 
CL J1059.2$-$1253      & 22&  \phantom{11}401$^{\scriptscriptstyle{+ 71}}_{\scriptscriptstyle{- 54}}$ & 19&  \phantom{11}613$^{\scriptscriptstyle{+ 81}}_{\scriptscriptstyle{- 75}}$\\ 
CL J1103.7$-$1245a     &  7&  \phantom{11}341$^{\scriptscriptstyle{+163}}_{\scriptscriptstyle{- 97}}$ &  7&  \phantom{11}371$^{\scriptscriptstyle{+ 82}}_{\scriptscriptstyle{-187}}$\\
CL J1103.7$-$1245      &  5&  \phantom{11}571$^{\scriptscriptstyle{+273}}_{\scriptscriptstyle{-162}}$ &  9&  \phantom{11}227$^{\scriptscriptstyle{+100}}_{\scriptscriptstyle{- 36}}$\\ 
RX J1117.4+0743        & 19& \phantom{1}1066$^{\scriptscriptstyle{+441}}_{\scriptscriptstyle{-227}}$ & 18& \phantom{1}1654$^{\scriptscriptstyle{+248}}_{\scriptscriptstyle{-156}}$\\
CL J1138.2$-$1133a     &  7&  \phantom{11}418$^{\scriptscriptstyle{+131}}_{\scriptscriptstyle{- 50}}$ &  7&  \phantom{11}486$^{\scriptscriptstyle{+657}}_{\scriptscriptstyle{- 57}}$\\ 
CL J1138.2$-$1133      & 20&  \phantom{11}484$^{\scriptscriptstyle{+123}}_{\scriptscriptstyle{- 79}}$ & 29&  \phantom{11}867$^{\scriptscriptstyle{+133}}_{\scriptscriptstyle{- 40}}$\\
CL J1216.8$-$1201      & 35&  \phantom{11}976$^{\scriptscriptstyle{+122}}_{\scriptscriptstyle{- 78}}$ & 31&  \phantom{11}959$^{\scriptscriptstyle{+117}}_{\scriptscriptstyle{- 71}}$\\ 
RX J1226.9+3332        & 43&  \phantom{11}926$^{\scriptscriptstyle{+127}}_{\scriptscriptstyle{-101}}$ &  7& \phantom{1}1199$^{\scriptscriptstyle{+698}}_{\scriptscriptstyle{-258}}$\\ 
XMMU J1230.3+1339      &  8&  \phantom{11}768$^{\scriptscriptstyle{+192}}_{\scriptscriptstyle{- 71}}$ &  5&  \phantom{11}605$^{\scriptscriptstyle{+279}}_{\scriptscriptstyle{- 60}}$\\
CL J1232.5$-$1250      & 26& \phantom{1}1025$^{\scriptscriptstyle{+235}}_{\scriptscriptstyle{- 91}}$ & 28& \phantom{1}1062$^{\scriptscriptstyle{+147}}_{\scriptscriptstyle{-129}}$\\ 
RDCS J1252$-$2927      & 21&  \phantom{11}787$^{\scriptscriptstyle{+110}}_{\scriptscriptstyle{-108}}$ & 16&  \phantom{11}775$^{\scriptscriptstyle{+143}}_{\scriptscriptstyle{- 92}}$\\ 
CL J1301.7$-$1139a     & 12&  \phantom{11}331$^{\scriptscriptstyle{+ 89}}_{\scriptscriptstyle{- 52}}$ &  5&  \phantom{11}586$^{\scriptscriptstyle{+176}}_{\scriptscriptstyle{-193}}$\\       
CL J1301.7$-$1139      & 20&  \phantom{11}543$^{\scriptscriptstyle{+187}}_{\scriptscriptstyle{- 81}}$ & 17&  \phantom{11}778$^{\scriptscriptstyle{+143}}_{\scriptscriptstyle{- 65}}$\\
CL J1324+3011          & 25&  \phantom{11}723$^{\scriptscriptstyle{+128}}_{\scriptscriptstyle{-102}}$ & 19& \phantom{1}1028$^{\scriptscriptstyle{+248}}_{\scriptscriptstyle{-203}}$\\ 
CL J1353.0$-$1137      &  8&  \phantom{11}277$^{\scriptscriptstyle{+ 62}}_{\scriptscriptstyle{- 19}}$ & 13&  \phantom{11}810$^{\scriptscriptstyle{+182}}_{\scriptscriptstyle{- 96}}$\\                  	  
CL J1354.2$-$1230      & 11&  \phantom{11}455$^{\scriptscriptstyle{+ 91}}_{\scriptscriptstyle{- 63}}$ & 10&  \phantom{11}446$^{\scriptscriptstyle{+220}}_{\scriptscriptstyle{-103}}$\\   
CL J1411.1$-$1148      & 14&  \phantom{11}665$^{\scriptscriptstyle{+112}}_{\scriptscriptstyle{- 73}}$ & 11&  \phantom{11}741$^{\scriptscriptstyle{+264}}_{\scriptscriptstyle{-135}}$\\
3C 295                 & 12& \phantom{1}1354$^{\scriptscriptstyle{+168}}_{\scriptscriptstyle{-228}}$ & 13& \phantom{1}1686$^{\scriptscriptstyle{+310}}_{\scriptscriptstyle{-190}}$\\
CL 1601+4253           & 41&  \phantom{11}727$^{\scriptscriptstyle{+108}}_{\scriptscriptstyle{- 97}}$ &  9&  \phantom{11}764$^{\scriptscriptstyle{+169}}_{\scriptscriptstyle{-107}}$\\
CL J1604+4304          & 10&  \phantom{11}500$^{\scriptscriptstyle{+110}}_{\scriptscriptstyle{-125}}$ &  5& \phantom{1}1158$^{\scriptscriptstyle{+339}}_{\scriptscriptstyle{- 84}}$\\             		  
CL J1604+4321          & 11&  \phantom{11}591$^{\scriptscriptstyle{+188}}_{\scriptscriptstyle{-162}}$ & 26&  \phantom{11}722$^{\scriptscriptstyle{+289}}_{\scriptscriptstyle{-144}}$\\          	  
MS 1621.5+2640         & 63&  \phantom{11}731$^{\scriptscriptstyle{+ 80}}_{\scriptscriptstyle{- 68}}$ & 34&  \phantom{11}780$^{\scriptscriptstyle{+ 83}}_{\scriptscriptstyle{- 51}}$\\                    	  
RX J1716.6+6708        & 18& \phantom{1}1288$^{\scriptscriptstyle{+237}}_{\scriptscriptstyle{-166}}$ & 13& \phantom{1}1261$^{\scriptscriptstyle{+162}}_{\scriptscriptstyle{-454}}$\\                        
XMMXCS J2215.9$-$1738  & 12&  \phantom{11}713$^{\scriptscriptstyle{+156}}_{\scriptscriptstyle{-215}}$ & 24&  \phantom{11}647$^{\scriptscriptstyle{+110}}_{\scriptscriptstyle{- 82}}$\\              		  
XMMU J2235.3$-$2557    & 17&  \phantom{11}851$^{\scriptscriptstyle{+218}}_{\scriptscriptstyle{-103}}$ & 10&  \phantom{11}698$^{\scriptscriptstyle{+148}}_{\scriptscriptstyle{- 98}}$\\   
\noalign{\smallskip}                  
\hline                               
\noalign{\smallskip}
\end{tabular}
$$ 
\end{table}

\section{Analysis and results}
\label{vsegre}

\subsection{Galaxy color segregation in velocity space}
\label{vcs}

To investigate the relative kinematics of red and blue galaxy
populations, we applied a set of tests.  As for the individual
clusters, we checked for different values of velocity dispersions of
the two galaxy populations by applying the standard F--test
(\citealt{Press1992}).  We found that only 6 (9) of the 41 clusters
show evidence of a kinematical difference at the 95\% (90\%) c.l.. In
most of these cases, that is 5 (7) clusters, we found $\sigma_{V,\rm
  blue}>\sigma_{V,\rm red}$ (see Table~\ref{table5}).

\begin{table}
\caption[]{Clusters where $\sigma_{V,\rm red} \ne \sigma_{V,\rm blue}$.}
         \label{table5}
                $$
         \begin{tabular}{l r}
\hline
\noalign{\smallskip}
\hline
\noalign{\smallskip}
\multicolumn{1}{l}{Cluster name}&\multicolumn{1}{c}{F--test}\\
\multicolumn{1}{l}{}&\multicolumn{1}{c}{Prob$_{\ne}$}\\
\hline
\noalign{\smallskip}
RX J0152.7$-$1357  & 93\%\\
CL 0303+1706$^{a}$         & 99.9\%\\
CL J1054.4$-$1146   & 98\%\\
CL J1059.2$-$1253   & 94\%\\
CL J1103.7$-$1245$^{a}$   & 97\%\\
RX J1117.4+0743    & 93\%\\
CL J1138.2$-$1133   & 99\%\\
CL J1353.0$-$1137   & 99\%\\
CL J1604+4304       & 97\%\\
\noalign{\smallskip}                  
\hline                               
\noalign{\smallskip}
\end{tabular}
$$ \tablefoot{$^{(a)}$ Cases with $\sigma_{V,\rm red} > \sigma_{V,\rm blue}$.
In other cases $\sigma_{V,\rm blue} > \sigma_{V,\rm red}$.}
\end{table}

\begin{figure}
\centering
\resizebox{\hsize}{!}{\includegraphics{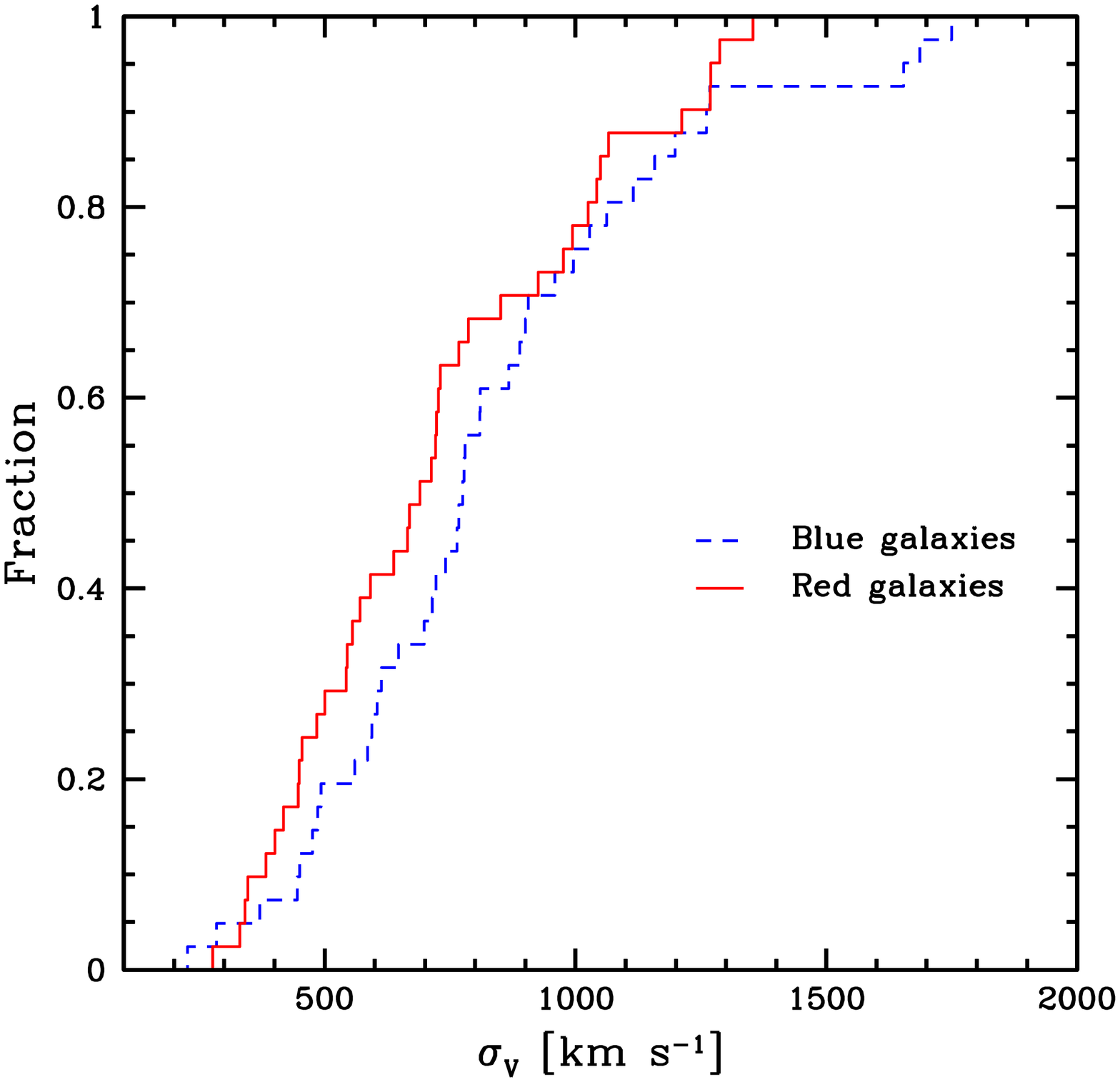}}
\caption
{Cumulative distributions of the velocity dispersions of red and blue
  galaxies.  }
\label{figKS}
\end{figure}

We compared the $\sigma_{V,\rm red}$ distribution and the
$\sigma_{V,\rm blue}$ distribution.  According to the KS--test, the
probability that the two distributions are drawn from the same parent
distribution is $\sim 24\%$, i.e. there is no evidence of a
significant difference. The cumulative distributions are compared in
Fig.~\ref{figKS} and the separation between the two distribution
median values is 75 \kss, that is $\sim 11\%$ of the median value of
global $\sigma_{V}$. The availability of two measures, $\sigma_{V,\rm
  red}$ and $\sigma_{V,\rm blue}$, for each cluster, allows us to also
apply the S-- and W--tests, which are more sensitive tests than the
KS--test, and to look for a possible systematic, even if small
difference. According to the S-- and W--tests, $\sigma_{V,\rm blue}$
is larger than $\sigma_{V,\rm red}$ at the $97.02\%$ and $99.45\%$
c.ls., respectively.  Out of 41 clusters, the number of those with
$\sigma_{V,\rm blue}>\sigma_{V,\rm red}$ is 27 (see inset in
Fig.~\ref{figss}).

\begin{figure}
\centering
\resizebox{\hsize}{!}{\includegraphics{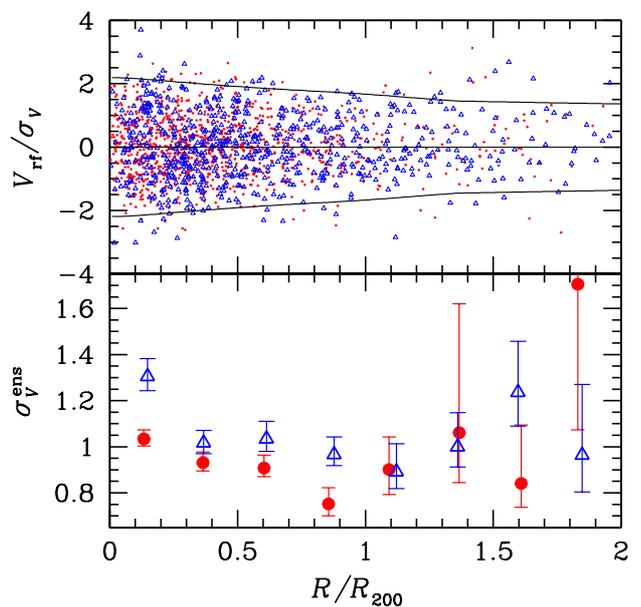}}
\caption
{{\em Upper panel}: rest-frame LOS velocity vs. projected
  clustercentric distance for the galaxies of the ensemble cluster.
  Small red dots indicate red galaxies and blue triangles indicate
  blue galaxies.  Black curves show the limits due to the escape
  velocity assuming a NFW mass profile (see text).  {\em Lower panel}:
  velocity dispersion profiles (VDPs) for red and blue galaxies of the
  ensemble cluster (solid red circles and blue open triangles). Data
  are binned in intervals of 0.25 Mpc in the 0--$2R_{200}$ range.
  The point abscissae are set to the averages of the $R/R_{200}$
    values of each sample within the bins. The error bars are
    determined via a bootstrap resampling procedure.}
\label{figvd}
\end{figure}

As a final test, we considered the projected phase space, that is the
rest-frame velocities $V_{\rm rf}=(V-<V>)/(1+z)$ vs clustercentric
distance $R$, of two ensemble clusters, one for red and one for blue
galaxies.  These two ensemble cluster data are obtained by stacking
all together the red (or blue) galaxies of each cluster normalizing
the velocity and the clustercentric distance of each galaxy with the
$\sigma_V$ and $R_{200}$ values of the parent cluster.  The result is
shown in Fig.~\ref{figvd} (upper panel), where we also trace the
limits due to the escape velocity in the cluster, assuming a typical
cluster mass distribution described by a NFW density profile with a
concentration parameter c=3.8, which is typical for halos of mass
$M=3$\mqua at $z=0.6$, the median values in our sample
(\citealt{Dolag2004}). The "trumpet shape" of the projected phase
  space data distribution and the good agreement with the escape
  velocity curves should be considered as a posteriori sanity check of
  the member selection procedure, which we made completely
  model-independent. In principle, the projection of possible non-member
  galaxies, likely blue field galaxies, onto the "trumpet shape"
  cannot be excluded.  However, according to the analysis of
  N-body cosmological simulations (\citealt{Biviano2006}), their
  effect should be that of slightly decreasing the value of the
  velocity dispersion, that is an opposite effect with respect to the
  segregation effect reported in the present and previous studies.

The respective VDPs for red and blue galaxies are shown in the lower
panel of Fig.~\ref{figvd}. The VDPs are shown to decline, as expected,
at least out to $\gtrsim R_{200}$. In the outer regions, the
uncertainties are very large and the fraction of possible interlopers,
that is galaxies outside the theoretical escape velocity curves,
survived to our member selection procedure, increases (see the upper
panel of Fig.~\ref{figvd}).  Within $R_{200}$ the VDP of blue galaxies
is higher than that of red galaxies and the difference is significant
at the $>99.99\%$ c.l. according to the $\chi^2$--test applied to the
values of the four bins, which combine a total of 936 red and 532 blue
galaxies.  Table~\ref{table6} summarizes the results of all the tests
we applied.

\begin{table*}
\caption[]{Statistical results about the relative kinematics of red and blue populations.}
         \label{table6}
                $$
         \begin{tabular}{l c c r c c c c r r r}
\hline
\noalign{\smallskip}
\hline
\noalign{\smallskip}
\multicolumn{1}{l}{Sample}
&\multicolumn{1}{c}{$z$ range}
&\multicolumn{1}{c}{$N_{\rm cl}$}
&\multicolumn{1}{c}{$N_{\rm g}$}
&\multicolumn{1}{c}{$N_{\rm cl,F}$}
&\multicolumn{1}{c}{$\Delta \sigma_{V}$}
&\multicolumn{1}{c}{$\Delta \sigma_{V}/\sigma_{V}$}
&\multicolumn{1}{c}{$N_{{\rm cl},{\rm b} > {\rm r}}$}
&\multicolumn{1}{c}{S--test}
&\multicolumn{1}{c}{W--test}
&\multicolumn{1}{c}{VDP $\chi^{2}$--test}
\\
\multicolumn{1}{l}{}
&\multicolumn{1}{c}{}
&\multicolumn{1}{c}{}
&\multicolumn{1}{c}{}
&\multicolumn{1}{c}{}
&\multicolumn{1}{c}{(km s$^{-1}$)}
&\multicolumn{1}{c}{}
&\multicolumn{1}{c}{}
&\multicolumn{1}{c}{Prob$_{\ne}$}
&\multicolumn{1}{c}{Prob$_{\ne}$}
&\multicolumn{1}{c}{Prob$_{\ne}$}
\\
\hline
\noalign{\smallskip}
Whole           & 0.4$\le z \le$1.5 & 41 &1674  &5 (7)& \phantom{1}75  &           11\%&       27 &       97\%&   99\%&>99.99\%\\
Low-z            & 0.4$\le z<$0.5    & 15 & 623  &2 (4)&           170  &           24\%&             12 & 98\%&  98\%& 99\%\\
Medium-z         & 0.5$\le z<$0.8    & 13 & 455  &2 (2)&           148  &           14\%&   \phantom{1}9 & 87\%&  93\%& 98\%\\
High-z           & 0.8$\le z \le$1.5 & 13 & 596  &1 (1)&  \phantom{1}48 & \phantom{1}6\%&   \phantom{1}6 & 50\%&  73\%& 96\%\\
\noalign{\smallskip}                  
\hline                               
\noalign{\smallskip}
\end{tabular}
$$ \tablefoot{ Column 1: sample name; Col.~2: the redshift range of
           the sample; Col.~3: the number of clusters, $N_{\rm cl}$;
           Col.~4: the number of galaxies in the sample, $N_{\rm g}$;
           Col.~5: the number of clusters with $\sigma_{V,\rm
             blue}>\sigma_{V,\rm red}$, with a 95\% (90\%) c.l.
           significant difference according to the F--test, $N_{\rm
             cl,F}$; Col.~6: the difference between the median values
           of the $\sigma_V$ distributions of red and blue galaxies,
           $\Delta \sigma_{V}$; Col.~7: the same, but normalized to
           the median value of $\sigma_{V}$ as computed using all
           galaxies, $\Delta \sigma_{V}/\sigma_{V}$; Col.~8: the
           number of clusters with $\sigma_{V,\rm blue}>\sigma_{V,\rm
             red}$, $N_{{\rm cl},{\rm b} > {\rm r}}$; Cols.~9 and 10:
           the probability of difference according to the S-- and W--tests;
Col.~11: the probability of difference according to
           the $\chi^{2}$--test applied to the VDPs of the ensemble
           cluster.  The results of the KS--test applied to the
           velocity dispersion distributions of red and blue galaxies
           are not listed since no evidence of significant difference
           is detected.  }
\end{table*}

\begin{figure}
\centering
\resizebox{\hsize}{!}{\includegraphics{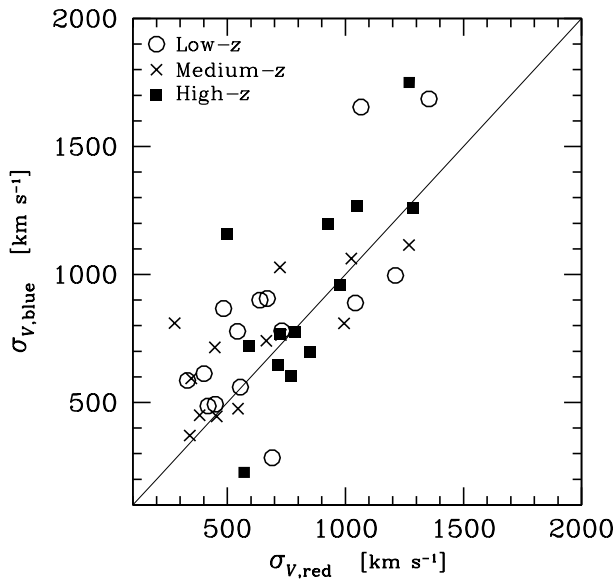}}
\caption
{Velocity dispersion of the blue galaxy population vs velocity
  dispersion of the red galaxy population for the three subsamples:
Low-$z$ (open circles), Medium-$z$ (crosses), and High-$z$ (solid squares).
}
\label{figS}
\end{figure}

\begin{figure}
\centering
\resizebox{\hsize}{!}{\includegraphics{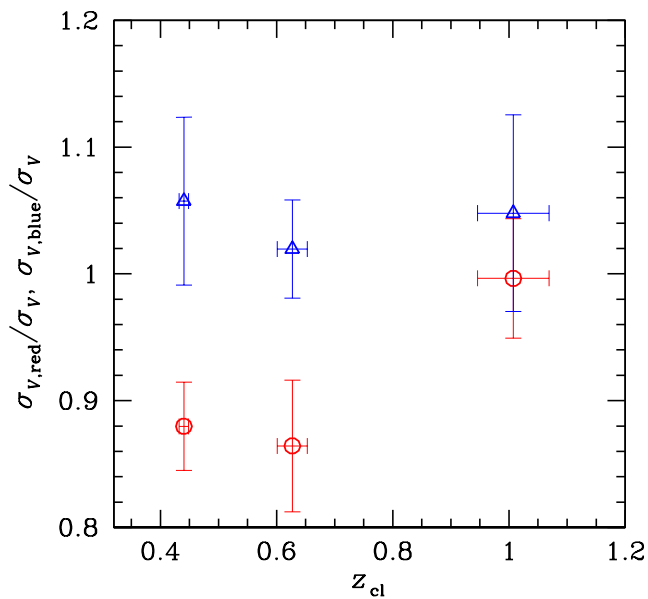}}
\caption
{Normalized velocity dispersion of the red galaxy population vs
  normalized velocity dispersion of the blue galaxy population for the
  three subsamples: Low-$z$, Medium-$z$ and High-$z$.}
\label{figvalorimedi}
\end{figure}

\begin{figure}
\centering
\resizebox{\hsize}{!}{\includegraphics{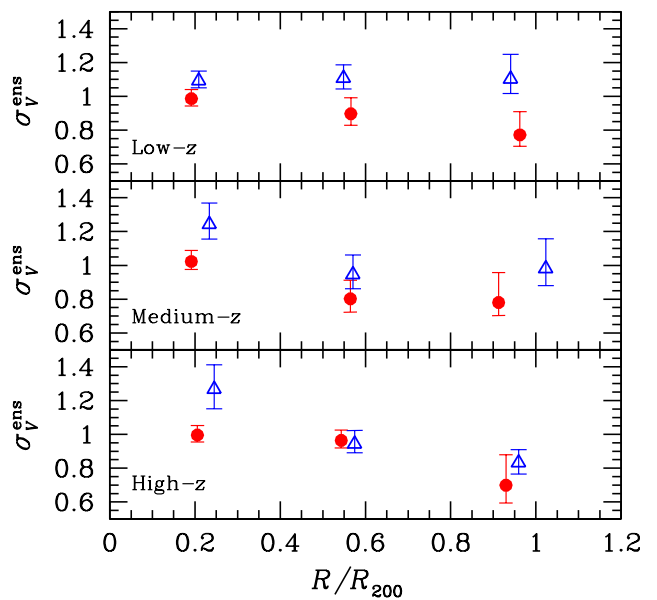}}
\caption
{Same as in Fig.~\ref{figvd}, but for the ensemble clusters of the
  Low-$z$, Medium-$z$, and High-$z$ subsamples ({\em top}, {\em medium}, and
  {\em bottom panels}).  Data are binned in intervals of 0.4 Mpc in
  the 0--$1.2R_{200}$ range.}
\label{figvdlowhigh}
\end{figure}

To investigate the variation of the kinematical difference between red
and blue galaxy populations at different redshifts, we divided the cluster
sample in three subsamples, Low-$z$ sample with $z<0.5$, Medium-$z$ with
$0.5\le z<0.8$, High-$z$ with $z\ge0.8$, having a roughly comparable
number of clusters and galaxies. We applied the above-described set of
tests to the three subsamples.  Fig.~\ref{figS} points out the
relation between $\sigma_{V,\rm blue}$ and $\sigma_{V,\rm red}$
separately for the three subsamples and shows that clusters of the
High-$z$ sample are equally split by $\sigma_{V,\rm blue} > \sigma_{V,\rm
  red}$ and $\sigma_{V,\rm blue} < \sigma_{V,\rm red}$ values.
Fig.~\ref{figvalorimedi} shows the variation of the normalized
$\sigma_{V,\rm blue}$ (and $\sigma_{V,\rm red}$) values with redshift,
no difference is found for the High-$z$ sample.  
To apply the VDP $\chi^2$--test, we considered data binned in three
intervals within $1.2R_{200}$ (see Fig.~\ref{figvdlowhigh}) for
a total of 349 red and 224 blue galaxies in the Low-$z$ sample, 247 red
and 160 blue galaxies in the Medium-$z$ sample, 369 red and 193 blue
galaxies in the High-$z$ sample.  The results of the whole set of tests
are listed in Table~\ref{table6}. We find that in the High-$z$ sample
there is no or poorer evidence of kinematical segregation with respect
to the other two samples, in particular with respect to the Low-$z$
sample.

\subsection{Galaxy luminosity segregation in velocity space}
\label{vls}

\begin{figure}
\centering
\resizebox{\hsize}{!}{\includegraphics{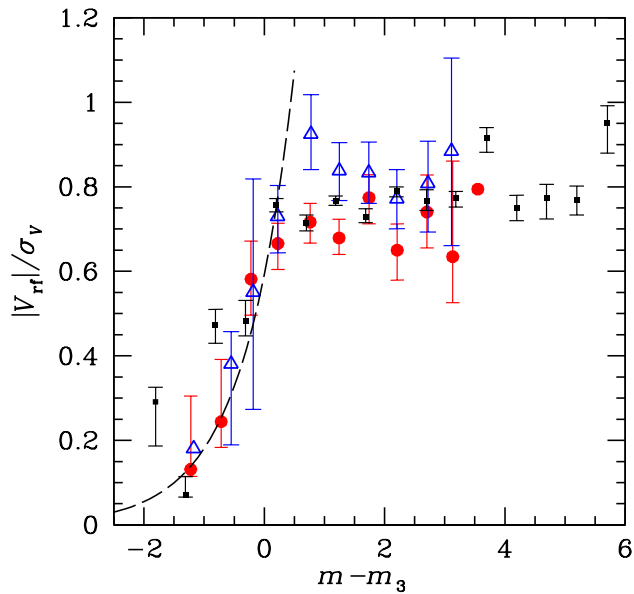}}
\caption
{Normalized velocities of red and blue galaxies (solid red circles and
  blue open triangles) vs $m-m_3$, where $m_3$ is the magnitude
    of the third brightest galaxy in each cluster. Data are binned in
  intervals of 0.5 mag. The error bars are obtained via a bootstrap
  resampling procedure. Points without error bars indicate
the values based on only three galaxies. The dashed line represents our fit
  for red galaxies in the $m-m_3\le 0.5$ region. The results of
  \citet{Biviano1992} are shown for comparison (small black squares).}
\label{figVLS}
\end{figure}

As in previous studies of VLS, our analysis is based on the
ensemble cluster. We restricted our analysis to galaxies within
$R_{200}$.  Because of the non homogeneity of the photometric data
among different clusters in our sample, we adopted the same approach
of \citet{Biviano1992} and normalized the magnitude of each galaxy
with the magnitude of the third brightest galaxy ($m_3$).  We used one
of the magnitude bands listed in Table~\ref{table1} preferring red or
NIR bands.  We analyzed the behavior of $|V_{\rm rf}|$ vs $m-m_3$ and
Figure~\ref{figVLS} highlights the main results: i) the red galaxies
have lower $|{V_{\rm rf}}|$ with respect to the blue galaxies,
independent of their magnitudes; ii) both red and blue galaxies show
evidence of velocity segregation.  To statistically evaluate VLS, we
also considered the correlation between $|V_{\rm rf}|/\sigma_V$ and $m-m_3$ for
galaxies having $m-m_3\le0.5$ or $m-m_3> 0.5$, where $m-m_3=0.5$ mag
is the threshold value suggested by the inspection of
Fig.~\ref{figVLS}. For the red (blue) galaxies with $m-m_3 \le 0.5$ we
find that $|V_{\rm rf}|$ and $m-m_3$ correlate at the $\sim 94\%$
c.l. ($\sim 90\%$) according to the Spearman test. No significant
correlation is found for red and blue galaxies with $m-m_3 > 0.5$.

\begin{figure}
\centering
\resizebox{\hsize}{!}{\includegraphics{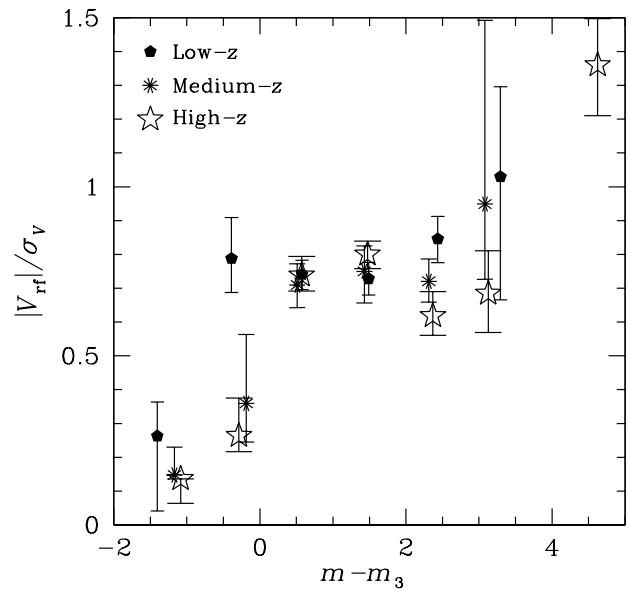}}
\caption
{Same as in Fig.~\ref{figVLS} but for all red+blue galaxies and
  analyzing Low-$z$, Medium-$z$, and High-$z$ subsamples separately (polygons,
  asterisks, and stars, resp.) }
\label{figVLSz}
\end{figure}

To investigate a possible dependence of VLS with redshift, we also
show the results for the three redshift subsamples. Although current
data are not sufficient to obtain firm conclusions, the visual
inspection of Fig.~\ref{figVLSz} suggests that the threshold value of
segregation in the Low-$z$ sample lies at brighter magnitudes than the
values in the Medium-$z$ and High-$z$ samples.

\section{Discussion}
\label{disc}

We find evidence of VCS in our sample of clusters.  In our analysis,
VCS is detected over the whole sampled galaxy luminosity range
(see Fig~\ref{figVLS}).  In particular, analyzing the clusters in
three redshift ranges separately, Table~\ref{table6} shows that the
amount of VCS decreases with increasing redshift and that very poor or
no evidence of segregation is found in the High-$z$
sample. Qualitatively, this is in agreement with the results of
\citet{Biviano2009} in EDisCS clusters at $z\sim 0.4$-0.8 (our
Low-$z$+Medium-$z$ samples), for which there is much less evidence of
VCS when compared to local clusters.  \citet{Crawford2014} found no
evidence of VCS between red sequence and blue cloud galaxies in five
distant cluster ($0.5<z<0.9$), but they did not analyze a local sample
for comparison.

We confirm that the effect of VCS is quantitatively small and,
therefore, difficult to detect.  For instance, according to the
KS--test, the $\sigma_V$ distributions of red and blue galaxies might
derive from the same parent distribution. We had to resort to very
sensitive tests such as the S-- and W--tests to detect a significant
difference in the $\sigma_V$ distributions.  Indeed, as most previous
positive detections in the literature, our most significant detection
is obtained stacking galaxies of all clusters.  Quantitatively, the
estimates of the difference in $\sigma_V$ reported in the literature
are of the order of 10-20\%, 30\% at most (\citealt{Biviano1997};
\citealt{Carlberg1997}; \citealt{Adami1998}; \citealt{deTheije1999};
\citealt{Haines2015}) and vary according to the selection of the two
populations. In fact, when the population is selected according to
higher values of star formation rate, its $\sigma_V$ is higher
(\citealt{Haines2015}).  The values we obtained in our study are
comparable with those reported in the literature. We estimated that
$\sigma_V$ of red and blue galaxies differ for $\Delta
\sigma_V/\sigma_V \sim$ 20\% and 10\% in the Low-$z$ and Medium-$z$
samples, and the median value of the ratios of the VDP binned values
for blue and red galaxies is 1.2 (within $R_{200}$, see
Fig.~\ref{figvd}).

Taking into account the difficulties related to the detection and
measure of VCS, any result is more reliable when obtained through an
homogeneous analysis rather than comparing results from different
authors. In this context, it is interesting that our results are in
line with that of \citet{Biviano2009} on a decreasing amount of VCS at
higher redshifts, although their claim holds for clusters at $z \sim
0.4$-0.8, while ours for clusters in the $0.8\le z\le 1.5$ range. The
poor evidence of VCS in high redshift clusters can be explained in a
scenario where the segregation develops as the result of a continuous,
regular smooth accretion of field blue galaxies, then possibly
evolving in red galaxies, into the cluster. VCS is erased when
cluster-cluster mergers drive violent relaxation, and the frequency of
such mergers is expected to be higher at higher redshift.  

Since our cluster sample spans a large range of masses, of about two
orders of magnitudes (see Table \ref{table2}) we also checked for a
possible dependence of VCS with mass.  We find no significant
correlation of $\sigma_{V,\rm blue}/\sigma_{V,\rm red}$ vs the cluster
mass.  This agrees with the fact that the evidence of VCS, which is
well known in clusters, is detected also in groups
(\citealt{Girardi2003}; \citealt{Ribeiro2010}).

To our knowledge, this is the first study where VLS is
detected in non local clusters. The $|V_{\rm rf}|$ vs $m-m_3$ relation
here detected is similar to that originally shown  by
\citet{Biviano1992}. Assuming the mass-follows-light hypothesis, the
faint galaxies can be described by a regime of velocity equipartition,
while the bright galaxies can be better described by a regime of
energy equipartition, with more massive objects being slower.  Since
we used red or NIR magnitude bands, the luminosity is a good indicator
of the stellar mass, but the mass-follows-light assumption is needed
to extrapolate our interpretation to the whole, halo galaxy mass.  As
already discussed by other authors (\citealt{Biviano1992};
\citealt{Mahajan2011}) the most likely cause for the observed VLS is
the dynamical friction process, whose characteristic time-scale
is inversely proportional to mass. 

We confirm that VLS holds for red galaxies and find for the first time
evidence that the same segregation also applies to the population of
blue galaxies.  In fact, both \citet{Adami1998} and
\citet{Ribeiro2013} found VLS for ellipticals/passive galaxies, but no
(or even opposite) effect is reported for galaxies of other types.
However, the inspection of Fig.~2b of \citet{Adami1998} suggests that
their non detection might be rather due to the large uncertainties
involved. The presence of the VLS for blue galaxies indicates that the
kinematical relaxation time-scale is shorter than the transformation
time-scale or that massive blue galaxies are robust against
environmental effects and possible transformation to S0 (e.g.,
\citealt{Moore1996}; \citealt{Bekki2011}). 

A more detailed comparison of our results with those of Biviano et al.
(\citeyear{Biviano1992}, see also our Fig.~\ref{figVLS}) shows two
differences. As a first difference, our $|V_{\rm rf}|$ and $m-m_3$
relation is steeper than their relation. The fit of the logarithm of
$|V_{\rm rf}|$ vs $m-m_3$ gives a slope of $0.52\pm0.05$ for red
galaxies only, and $0.48\pm0.04$ for all galaxies.  For comparison,
\citet{Biviano1992} obtained a value of 0.2, which is that expected in
the case of energy equipartition (assuming a constant mass-to-light
ratio).  \citet{Adami1998} also claim that a 0.2 slope is consistent
with their results, but large uncertainties are shown in their Fig.~1a
and the inspection of their Fig.~2a for ellipticals rather suggests a
steeper slope.  The second difference concerns the threshold value
between the two kinematic regimes.  \citet{Biviano1992} indicate a
threshold value around $m_3$ and \citet{Adami1998} report that
  VLS concerns about four galaxies per cluster. In the whole
  sample, our results rather suggest $m_3+0.5$ (Fig.~\ref{figVLS}) and
  we found that VLS concerns seven galaxies per cluster, computed
  as the median value of the numbers of galaxies with $m-m_3 \le 0.5$
  in each cluster. However, both \citet{Biviano1992} and
  \citet{Adami1998} analyze local clusters and the inspection of
  Fig.~\ref{figVLSz} indicates that the threshold value of VLS in
  the Low-$z$ sample lies at brighter magnitudes than the values in
  the Medium-$z$ and High-$z$ samples. Present data do not allow a
  precise quantitative conclusion, but suggest that the dipendence of
  the segregation threshold should be taken into account.  In
  particular, the possible explanation for a fainter threshold at
  higher redshifts might be that clusters at higher redshift have
higher density. Per given galaxy mass, the greater the density of the
surrounding medium, the stronger the effect of dynamical friction.

\section{Summary and conclusions}
\label{summ}

In this study we present our results about color and luminosity
segregation in velocity space (VCS and VLS, resp.) for a sample of 41
clusters at intermediate and high redshifts ($0.4 \lesssim z \lesssim
1.5$), for a total of 4172 galaxies. The data have been taken from
different sources in the literature, with the constraint that data for
each single cluster come from one single source. Moreover, we applied
homogeneous preliminary procedures to select cluster members, compute
global cluster properties, in particular the LOS velocity dispersions
$\sigma_{V}$, and separate blue from red galaxies. We restricted our
analysis to the 1674 member galaxies within 2$R_{200}$ with
photometric or spectroscopic information, 1023 red and 651 blue
galaxies.  We applied a set of different tests to study VCS and VLS.
We used both the estimates of velocity dispersion for each individual
cluster and the properties of an ensemble cluster obtained by stacking
together galaxies of many clusters.

The main results of our analysis are summarized as follows.

\begin{itemize}

\item From the analysis of the whole sample we detect evidence of VCS
  according to several tests (S--, W--, and VDP $\chi^2$--tests), with
  the blue galaxy population having a larger $\sigma_{V}$
than the red galaxy population.

\item When analyzing three subsamples at different redshifts (Low-$z$
  with $z<0.5$, Medium-$z$ with $0.5\le z<0.8$, High-$z$ with $z\ge0.8$),
  very poor or no evidence of VCS is found in the High-$z$ sample. The
  fact that VCS is weaker at higher redshifts has been already pointed
  out by \citet{Biviano2009}, although our threshold of no detection
  is at higher $z$ than theirs. The disappearance of the VCS for
  distant clusters can be
  explained when considering that our High-$z$ sample is very close to
  the epoch of cluster formation, with major mergers driving violent
  relaxation which leads to the velocity equipartition regime.

\item In agreement with previous studies we confirm that the effect of
  VCS is quantitatively small (10-20\% in the $\sigma_{V}$ estimate)
  and requires sensitive tests or the VDP analysis based on many
  galaxies.  We conclude that VCS is an elusive effect, that might
  partly explains the discrepant claims reported in the literature on
  this issue.

\item VCS concerns the whole magnitude range that we covered, $\sim 4$
  magnitudes down to $m_3$; more clusters are needed to sample the
  bright end to obtain firm conclusions.

\item We detect evidence of VLS for galaxies more luminous than
  $m_3+0.5$, brighter galaxies having lower velocities. Qualitatively,
  this result is similar to that found for local clusters, but we note
  and discuss minor differences, e.g. in the threshold value of the
  segregation.

\item VLS concerns both red and blue galaxies. The
  latter finding still not reported in the literature, not even for
  local clusters.

\end{itemize}

Finally, we note that there is a strong correlation between $\sigma_V$
based on red galaxies and $\sigma_V$ based on blue galaxies and, in
particular, we find no significant bias in the High-$z$
sample. Although the appropriate mass calibration has to be
determined, this result suggests that both red and blue galaxies can
be used as tracers of the cluster mass distribution out to high
redshift. This result has interesting implications for the
cosmological application of the velocity dispersion measurements that
the Euclid satellite will make possible by targeting H$\alpha$-emitting, star-forming
galaxies in its spectroscopic survey
\citep[e.g.][]{Laureijs2011,Sartoris2016}.

\begin{acknowledgements}
We thank the referee for useful comments.  We thank P. Rosati for
providing us data on XMMU\ J2235.3$-$2557.  M.A., and A.B. and
M.N. acknowledge financial support from PRIN-INAF 2014
1.05.01.94.02. M.G.  acknowledges financial support from the
University of Trieste through the program ``Finanziamento di Ateneo
per progetti di ricerca scientifica - FRA 2015''.  S.B. acknowledges
financial support from the PRIN-MIUR 201278X4FL grant and from the
``InDark'' INFN Grant. This research has made use of the NASA/IPAC
Extragalactic Database (NED) which is operated by the Jet Propulsion
Laboratory, California Institute of Technology, under contract with
the National Aeronautics and Space Administration.

\end{acknowledgements}

\bibliographystyle{aa}
\bibliography{biblio}

\begin{thebibliography}{91}
\expandafter\ifx\csname natexlab\endcsname\relax\def\natexlab#1{#1}\fi

\bibitem[{{Abraham} {et~al.}(1996){Abraham}, {Smecker-Hane}, {Hutchings},
  {Carlberg}, {Yee}, {Ellingson}, {Morris}, {Oke}, \& {Rigler}}]{Abraham1996}
{Abraham}, R.~G., {Smecker-Hane}, T.~A., {Hutchings}, J.~B., {et~al.} 1996,
  \apj, 471, 694

\bibitem[{{Adami} {et~al.}(1998){Adami}, {Biviano}, \& {Mazure}}]{Adami1998}
{Adami}, C., {Biviano}, A., \& {Mazure}, A. 1998, \aap, 331, 439

\bibitem[{{Ashman} {et~al.}(1994){Ashman}, {Bird}, \& {Zepf}}]{Ashman1994}
{Ashman}, K.~M., {Bird}, C.~M., \& {Zepf}, S.~E. 1994, \aj, 108, 2348

\bibitem[{{Beers} {et~al.}(1990){Beers}, {Flynn}, \& {Gebhardt}}]{Beers1990}
{Beers}, T.~C., {Flynn}, K., \& {Gebhardt}, K. 1990, \aj, 100, 32

\bibitem[{{Bekki} \& {Couch}(2011)}]{Bekki2011}
{Bekki}, K. \& {Couch}, W.~J. 2011, \mnras, 415, 1783

\bibitem[{{Biviano} {et~al.}(1996){Biviano}, {Durret}, {Gerbal}, {Le Fevre},
  {Lobo}, {Mazure}, \& {Slezak}}]{Biviano1996}
{Biviano}, A., {Durret}, F., {Gerbal}, D., {et~al.} 1996, \aap, 311, 95

\bibitem[{{Biviano} \& {Girardi}(2003)}]{Biviano2003}
{Biviano}, A. \& {Girardi}, M. 2003, \apj, 585, 205

\bibitem[{{Biviano} {et~al.}(1992){Biviano}, {Girardi}, {Giuricin},
  {Mardirossian}, \& {Mezzetti}}]{Biviano1992}
{Biviano}, A., {Girardi}, M., {Giuricin}, G., {Mardirossian}, F., \&
  {Mezzetti}, M. 1992, \apj, 396, 35

\bibitem[{{Biviano} \& {Katgert}(2004)}]{Biviano2004}
{Biviano}, A. \& {Katgert}, P. 2004, \aap, 424, 779

\bibitem[{{Biviano} {et~al.}(1997){Biviano}, {Katgert}, {Mazure}, {Moles}, {den
  Hartog}, {Perea}, \& {Focardi}}]{Biviano1997}
{Biviano}, A., {Katgert}, P., {Mazure}, A., {et~al.} 1997, \aap, 321, 84

\bibitem[{{Biviano} {et~al.}(2006){Biviano}, {Murante}, {Borgani}, {Diaferio},
  {Dolag}, \& {Girardi}}]{Biviano2006}
{Biviano}, A., {Murante}, G., {Borgani}, S., {et~al.} 2006, \aap, 456, 23

\bibitem[{{Biviano} \& {Poggianti}(2009)}]{Biviano2009}
{Biviano}, A. \& {Poggianti}, B.~M. 2009, \aap, 501, 419

\bibitem[{{Biviano} {et~al.}(2013){Biviano}, {Rosati}, {Balestra}, {Mercurio},
  {Girardi}, {Nonino}, {Grillo}, {Scodeggio}, {Lemze}, {Kelson}, {Umetsu},
  {Postman}, {Zitrin}, {Czoske}, {Ettori}, {Fritz}, {Lombardi}, {Maier},
  {Medezinski}, {Mei}, {Presotto}, {Strazzullo}, {Tozzi}, {Ziegler},
  {Annunziatella}, {Bartelmann}, {Benitez}, {Bradley}, {Brescia}, {Broadhurst},
  {Coe}, {Demarco}, {Donahue}, {Ford}, {Gobat}, {Graves}, {Koekemoer},
  {Kuchner}, {Melchior}, {Meneghetti}, {Merten}, {Moustakas}, {Munari}, {Reg{\H
  o}s}, {Sartoris}, {Seitz}, \& {Zheng}}]{Biviano2013}
{Biviano}, A., {Rosati}, P., {Balestra}, I., {et~al.} 2013, \aap, 558, A1

\bibitem[{{Borgani} {et~al.}(1997){Borgani}, {Gardini}, {Girardi}, \&
  {Gottl{\"o}ber}}]{Borgani1997}
{Borgani}, S., {Gardini}, A., {Girardi}, M., \& {Gottl{\"o}ber}, S. 1997, \na,
  2, 119

\bibitem[{{Caldwell} \& {Rose}(1997)}]{Caldwell1997}
{Caldwell}, N. \& {Rose}, J.~A. 1997, \aj, 113, 492

\bibitem[{{Carlberg} {et~al.}(1997){Carlberg}, {Yee}, {Ellingson}, {Morris},
  {Abraham}, {Gravel}, {Pritchet}, {Smecker-Hane}, {Hartwick}, {Hesser},
  {Hutchings}, \& {Oke}}]{Carlberg1997}
{Carlberg}, R.~G., {Yee}, H.~K.~C., {Ellingson}, E., {et~al.} 1997, \apjl, 476,
  L7

\bibitem[{{Carrasco} {et~al.}(2007){Carrasco}, {Cypriano}, {Neto}, {Cuevas},
  {Sodr{\'e}}, {de Oliveira}, \& {Ramirez}}]{Carrasco2007}
{Carrasco}, E.~R., {Cypriano}, E.~S., {Neto}, G.~B.~L., {et~al.} 2007, \apj,
  664, 777

\bibitem[{{Chincarini} \& {Rood}(1977)}]{Chincarini1977}
{Chincarini}, G. \& {Rood}, H.~J. 1977, \apj, 214, 351

\bibitem[{{Colless} \& {Dunn}(1996)}]{Colless1996}
{Colless}, M. \& {Dunn}, A.~M. 1996, \apj, 458, 435

\bibitem[{{Crawford} {et~al.}(2014){Crawford}, {Wirth}, \&
  {Bershady}}]{Crawford2014}
{Crawford}, S.~M., {Wirth}, G.~D., \& {Bershady}, M.~A. 2014, \apj, 786, 30

\bibitem[{{Danese} {et~al.}(1980){Danese}, {de Zotti}, \& {di
  Tullio}}]{Danese1980}
{Danese}, L., {de Zotti}, G., \& {di Tullio}, G. 1980, \aap, 82, 322

\bibitem[{{de Theije} \& {Katgert}(1999)}]{deTheije1999}
{de Theije}, P.~A.~M. \& {Katgert}, P. 1999, \aap, 341, 371

\bibitem[{{Demarco} {et~al.}(2010){Demarco}, {Gobat}, {Rosati}, {Lidman},
  {Rettura}, {Nonino}, {van der Wel}, {Jee}, {Blakeslee}, {Ford}, \&
  {Postman}}]{Demarco2010}
{Demarco}, R., {Gobat}, R., {Rosati}, P., {et~al.} 2010, \apj, 725, 1252

\bibitem[{{Demarco} {et~al.}(2007){Demarco}, {Rosati}, {Lidman}, {Girardi},
  {Nonino}, {Rettura}, {Strazzullo}, {van der Wel}, {Ford}, {Mainieri},
  {Holden}, {Stanford}, {Blakeslee}, {Gobat}, {Postman}, {Tozzi}, {Overzier},
  {Zirm}, {Ben{\'{\i}}tez}, {Homeier}, {Illingworth}, {Infante}, {Jee}, {Mei},
  {Menanteau}, {Motta}, {Zheng}, {Clampin}, \& {Hartig}}]{Demarco2007}
{Demarco}, R., {Rosati}, P., {Lidman}, C., {et~al.} 2007, \apj, 663, 164

\bibitem[{{Dolag} {et~al.}(2004){Dolag}, {Bartelmann}, {Perrotta},
  {Baccigalupi}, {Moscardini}, {Meneghetti}, \& {Tormen}}]{Dolag2004}
{Dolag}, K., {Bartelmann}, M., {Perrotta}, F., {et~al.} 2004, \aap, 416, 853

\bibitem[{{Dressler}(1980)}]{Dressler1980}
{Dressler}, A. 1980, \apj, 236, 351

\bibitem[{{Dressler} {et~al.}(1999){Dressler}, {Smail}, {Poggianti}, {Butcher},
  {Couch}, {Ellis}, \& {Oemler}}]{Dressler1999}
{Dressler}, A., {Smail}, I., {Poggianti}, B.~M., {et~al.} 1999, \apjs, 122, 51

\bibitem[{{Ellingson} {et~al.}(1998){Ellingson}, {Yee}, {Abraham}, {Morris}, \&
  {Carlberg}}]{Ellingson1998}
{Ellingson}, E., {Yee}, H.~K.~C., {Abraham}, R.~G., {Morris}, S.~L., \&
  {Carlberg}, R.~G. 1998, \apjs, 116, 247

\bibitem[{{Ellingson} {et~al.}(1997){Ellingson}, {Yee}, {Abraham}, {Morris},
  {Carlberg}, \& {Smecker-Hane}}]{Ellingson1997}
{Ellingson}, E., {Yee}, H.~K.~C., {Abraham}, R.~G., {et~al.} 1997, \apjs, 113,
  1

\bibitem[{{Fabricant} {et~al.}(1994){Fabricant}, {Bautz}, \&
  {McClintock}}]{Fabricant1994}
{Fabricant}, D.~G., {Bautz}, M.~W., \& {McClintock}, J.~E. 1994, \aj, 107, 8

\bibitem[{{Fadda} {et~al.}(1996){Fadda}, {Girardi}, {Giuricin}, {Mardirossian},
  \& {Mezzetti}}]{Fadda1996}
{Fadda}, D., {Girardi}, M., {Giuricin}, G., {Mardirossian}, F., \& {Mezzetti},
  M. 1996, \apj, 473, 670

\bibitem[{{Fassbender} {et~al.}(2011){Fassbender}, {B{\"o}hringer}, {Santos},
  {Pratt}, {{\v S}uhada}, {Kohnert}, {Lerchster}, {Rovilos}, {Pierini}, {Chon},
  {Schwope}, {Lamer}, {M{\"u}hlegger}, {Rosati}, {Quintana}, {Nastasi}, {de
  Hoon}, {Seitz}, \& {Mohr}}]{Fassbender2011}
{Fassbender}, R., {B{\"o}hringer}, H., {Santos}, J.~S., {et~al.} 2011, \aap,
  527, A78

\bibitem[{{Ferrari} {et~al.}(2005){Ferrari}, {Benoist}, {Maurogordato},
  {Cappi}, \& {Slezak}}]{Ferrari2005}
{Ferrari}, C., {Benoist}, C., {Maurogordato}, S., {Cappi}, A., \& {Slezak}, E.
  2005, \aap, 430, 19

\bibitem[{{Fukugita} {et~al.}(1995){Fukugita}, {Shimasaku}, \&
  {Ichikawa}}]{Fukugita1995}
{Fukugita}, M., {Shimasaku}, K., \& {Ichikawa}, T. 1995, \pasp, 107, 945

\bibitem[{{Gerken} {et~al.}(2004){Gerken}, {Ziegler}, {Balogh}, {Gilbank},
  {Fritz}, \& {J{\"a}ger}}]{Gerken2004}
{Gerken}, B., {Ziegler}, B., {Balogh}, M., {et~al.} 2004, \aap, 421, 59

\bibitem[{{Gioia} {et~al.}(1999){Gioia}, {Henry}, {Mullis}, {Ebeling}, \&
  {Wolter}}]{Gioia1999}
{Gioia}, I.~M., {Henry}, J.~P., {Mullis}, C.~R., {Ebeling}, H., \& {Wolter}, A.
  1999, \aj, 117, 2608

\bibitem[{{Girardi} {et~al.}(1993){Girardi}, {Biviano}, {Giuricin},
  {Mardirossian}, \& {Mezzetti}}]{Girardi1993}
{Girardi}, M., {Biviano}, A., {Giuricin}, G., {Mardirossian}, F., \&
  {Mezzetti}, M. 1993, \apj, 404, 38

\bibitem[{{Girardi} {et~al.}(1996){Girardi}, {Fadda}, {Giuricin},
  {Mardirossian}, {Mezzetti}, \& {Biviano}}]{Girardi1996}
{Girardi}, M., {Fadda}, D., {Giuricin}, G., {et~al.} 1996, \apj, 457, 61

\bibitem[{{Girardi} {et~al.}(2015){Girardi}, {Mercurio}, {Balestra}, {Nonino},
  {Biviano}, {Grillo}, {Rosati}, {Annunziatella}, {Demarco}, {Fritz}, {Gobat},
  {Lemze}, {Presotto}, {Scodeggio}, {Tozzi}, {Bartosch Caminha}, {Brescia},
  {Coe}, {Kelson}, {Koekemoer}, {Lombardi}, {Medezinski}, {Postman},
  {Sartoris}, {Umetsu}, {Zitrin}, {Boschin}, {Czoske}, {De Lucia}, {Kuchner},
  {Maier}, {Meneghetti}, {Monaco}, {Monna}, {Munari}, {Seitz}, {Verdugo}, \&
  {Ziegler}}]{Girardi2015}
{Girardi}, M., {Mercurio}, A., {Balestra}, I., {et~al.} 2015, \aap, 579, A4

\bibitem[{{Girardi} {et~al.}(2003){Girardi}, {Rigoni}, {Mardirossian}, \&
  {Mezzetti}}]{Girardi2003}
{Girardi}, M., {Rigoni}, E., {Mardirossian}, F., \& {Mezzetti}, M. 2003, \aap,
  406, 403

\bibitem[{{Goto}(2005)}]{Goto2005}
{Goto}, T. 2005, \mnras, 359, 1415

\bibitem[{{Haines} {et~al.}(2015){Haines}, {Pereira}, {Smith}, {Egami},
  {Babul}, {Finoguenov}, {Ziparo}, {McGee}, {Rawle}, {Okabe}, \&
  {Moran}}]{Haines2015}
{Haines}, C.~P., {Pereira}, M.~J., {Smith}, G.~P., {et~al.} 2015, \apj, 806,
  101

\bibitem[{{Halliday} {et~al.}(2004){Halliday}, {Milvang-Jensen}, {Poirier},
  {Poggianti}, {Jablonka}, {Arag{\'o}n-Salamanca}, {Saglia}, {De Lucia},
  {Pell{\'o}}, {Simard}, {Clowe}, {Rudnick}, {Dalcanton}, {White}, \&
  {Zaritsky}}]{Halliday2004}
{Halliday}, C., {Milvang-Jensen}, B., {Poirier}, S., {et~al.} 2004, \aap, 427,
  397

\bibitem[{{Hammer} {et~al.}(1997){Hammer}, {Flores}, {Lilly}, {Crampton}, {Le
  F{\`e}vre}, {Rola}, {Mallen-Ornelas}, {Schade}, \& {Tresse}}]{Hammer1997}
{Hammer}, F., {Flores}, H., {Lilly}, S.~J., {et~al.} 1997, \apj, 481, 49

\bibitem[{{Hern{\'a}ndez-Fern{\'a}ndez}
  {et~al.}(2014){Hern{\'a}ndez-Fern{\'a}ndez}, {Haines}, {Diaferio},
  {Iglesias-P{\'a}ramo}, {Mendes de Oliveira}, \& {Vilchez}}]{Hernandez2014}
{Hern{\'a}ndez-Fern{\'a}ndez}, J.~D., {Haines}, C.~P., {Diaferio}, A., {et~al.}
  2014, \mnras, 438, 2186

\bibitem[{{Hilton} {et~al.}(2010){Hilton}, {Lloyd-Davies}, {Stanford}, {Stott},
  {Collins}, {Romer}, {Hosmer}, {Hoyle}, {Kay}, {Liddle}, {Mehrtens}, {Miller},
  {Sahl{\'e}n}, \& {Viana}}]{Hilton2010}
{Hilton}, M., {Lloyd-Davies}, E., {Stanford}, S.~A., {et~al.} 2010, \apj, 718,
  133

\bibitem[{{Hwang} \& {Lee}(2008)}]{HwangLee2008}
{Hwang}, H.~S. \& {Lee}, M.~G. 2008, \apj, 676, 218

\bibitem[{{J{\o}rgensen} \& {Chiboucas}(2013)}]{Jorgensen2013}
{J{\o}rgensen}, I. \& {Chiboucas}, K. 2013, \aj, 145, 77

\bibitem[{{Katgert} {et~al.}(1996){Katgert}, {Mazure}, {Perea}, {den Hartog},
  {Moles}, {Le Fevre}, {Dubath}, {Focardi}, {Rhee}, {Jones}, {Escalera},
  {Biviano}, {Gerbal}, \& {Giuricin}}]{Katgert1996}
{Katgert}, P., {Mazure}, A., {Perea}, J., {et~al.} 1996, \aap, 310, 8

\bibitem[{{Laureijs} {et~al.}(2011){Laureijs}, {Amiaux}, {Arduini},
  {Augu{\`e}res}, {Brinchmann}, {Cole}, {Cropper}, {Dabin}, {Duvet}, {Ealet},
  \& et~al.}]{Laureijs2011}
{Laureijs}, R., {Amiaux}, J., {Arduini}, S., {et~al.} 2011, ArXiv e-prints
  [\eprint[arXiv]{1110.3193}]

\bibitem[{{Lederman}(1984)}]{Lederman1984}
{Lederman}, W. 1984, {Handbook of applicable mathematics. Vol.6,A: Statistics;
  Vol.6,B: Statistics} (Wiley-Interscience Publication, Chichester)

\bibitem[{{Lerchster} {et~al.}(2011){Lerchster}, {Seitz}, {Brimioulle},
  {Fassbender}, {Rovilos}, {B{\"o}hringer}, {Pierini}, {Kilbinger},
  {Finoguenov}, {Quintana}, \& {Bender}}]{Lerchster2011}
{Lerchster}, M., {Seitz}, S., {Brimioulle}, F., {et~al.} 2011, \mnras, 411,
  2667

\bibitem[{{Lubin} {et~al.}(2002){Lubin}, {Oke}, \& {Postman}}]{Lubin2002}
{Lubin}, L.~M., {Oke}, J.~B., \& {Postman}, M. 2002, \aj, 124, 1905

\bibitem[{{Mahajan} {et~al.}(2011){Mahajan}, {Mamon}, \&
  {Raychaudhury}}]{Mahajan2011}
{Mahajan}, S., {Mamon}, G.~A., \& {Raychaudhury}, S. 2011, \mnras, 416, 2882

\bibitem[{{Melnick} \& {Sargent}(1977)}]{Melnick1977}
{Melnick}, J. \& {Sargent}, W.~L.~W. 1977, \apj, 215, 401

\bibitem[{{Milvang-Jensen} {et~al.}(2008){Milvang-Jensen}, {Noll}, {Halliday},
  {Poggianti}, {Jablonka}, {Arag{\'o}n-Salamanca}, {Saglia}, {Nowak}, {von der
  Linden}, {De Lucia}, {Pell{\'o}}, {Moustakas}, {Poirier}, {Bamford}, {Clowe},
  {Dalcanton}, {Rudnick}, {Simard}, {White}, \& {Zaritsky}}]{Mil2008}
{Milvang-Jensen}, B., {Noll}, S., {Halliday}, C., {et~al.} 2008, \aap, 482, 419

\bibitem[{{Mohr} {et~al.}(1996){Mohr}, {Geller}, \& {Wegner}}]{Mohr1996}
{Mohr}, J.~J., {Geller}, M.~J., \& {Wegner}, G. 1996, \aj, 112, 1816

\bibitem[{{Moore} {et~al.}(1996){Moore}, {Katz}, {Lake}, {Dressler}, \&
  {Oemler}}]{Moore1996}
{Moore}, B., {Katz}, N., {Lake}, G., {Dressler}, A., \& {Oemler}, A. 1996,
  \nat, 379, 613

\bibitem[{{Moss} \& {Dickens}(1977)}]{Moss1977}
{Moss}, C. \& {Dickens}, R.~J. 1977, \mnras, 178, 701

\bibitem[{{Munari} {et~al.}(2013){Munari}, {Biviano}, {Borgani}, {Murante}, \&
  {Fabjan}}]{Munari2013}
{Munari}, E., {Biviano}, A., {Borgani}, S., {Murante}, G., \& {Fabjan}, D.
  2013, \mnras, 430, 2638

\bibitem[{{Owen} {et~al.}(2005){Owen}, {Ledlow}, {Keel}, {Wang}, \&
  {Morrison}}]{Owen2005}
{Owen}, F.~N., {Ledlow}, M.~J., {Keel}, W.~C., {Wang}, Q.~D., \& {Morrison},
  G.~E. 2005, \aj, 129, 31

\bibitem[{{Owers} {et~al.}(2011){Owers}, {Nulsen}, \& {Couch}}]{Owers2011}
{Owers}, M.~S., {Nulsen}, P.~E.~J., \& {Couch}, W.~J. 2011, \apj, 741, 122

\bibitem[{{Pisani}(1993)}]{Pisani1993}
{Pisani}, A. 1993, \mnras, 265, 706

\bibitem[{{Poggianti}(1997)}]{Poggianti1997}
{Poggianti}, B.~M. 1997, \aaps, 122 [\eprint{astro-ph/9608029}]

\bibitem[{{Postman} {et~al.}(1998){Postman}, {Lubin}, \& {Oke}}]{Postman1998}
{Postman}, M., {Lubin}, L.~M., \& {Oke}, J.~B. 1998, \aj, 116, 560

\bibitem[{{Postman} {et~al.}(2001){Postman}, {Lubin}, \& {Oke}}]{Postman2001}
{Postman}, M., {Lubin}, L.~M., \& {Oke}, J.~B. 2001, \aj, 122, 1125

\bibitem[{{Press} {et~al.}(1992){Press}, {Teukolsky}, {Vetterling}, \&
  {Flannery}}]{Press1992}
{Press}, W.~H., {Teukolsky}, S.~A., {Vetterling}, W.~T., \& {Flannery}, B.~P.
  1992, {Numerical recipes in FORTRAN. The art of scientific computing}
  (Cambridge University Press, Cambridge)

\bibitem[{{Ribeiro} {et~al.}(2013){Ribeiro}, {Lopes}, \&
  {Rembold}}]{Ribeiro2013}
{Ribeiro}, A.~L.~B., {Lopes}, P.~A.~A., \& {Rembold}, S.~B. 2013, \aap, 556,
  A74

\bibitem[{{Ribeiro} {et~al.}(2010){Ribeiro}, {Lopes}, \&
  {Trevisan}}]{Ribeiro2010}
{Ribeiro}, A.~L.~B., {Lopes}, P.~A.~A., \& {Trevisan}, M. 2010, \mnras, 409,
  L124

\bibitem[{{Rines} {et~al.}(2013){Rines}, {Geller}, {Diaferio}, \&
  {Kurtz}}]{Rines2013}
{Rines}, K., {Geller}, M.~J., {Diaferio}, A., \& {Kurtz}, M.~J. 2013, \apj,
  767, 15

\bibitem[{{Rines} {et~al.}(2003){Rines}, {Geller}, {Kurtz}, \&
  {Diaferio}}]{Rines2003}
{Rines}, K., {Geller}, M.~J., {Kurtz}, M.~J., \& {Diaferio}, A. 2003, \aj, 126,
  2152

\bibitem[{{Rines} {et~al.}(2005){Rines}, {Geller}, {Kurtz}, \&
  {Diaferio}}]{Rines2005}
{Rines}, K., {Geller}, M.~J., {Kurtz}, M.~J., \& {Diaferio}, A. 2005, \aj, 130,
  1482

\bibitem[{{Rosati} {et~al.}(2014){Rosati}, {Balestra}, {Grillo}, {Mercurio},
  {Nonino}, {Biviano}, {Girardi}, {Vanzella}, \& {Clash-VLT Team}}]{Rosati2014}
{Rosati}, P., {Balestra}, I., {Grillo}, C., {et~al.} 2014, The Messenger, 158,
  48

\bibitem[{{Rosati} {et~al.}(2009){Rosati}, {Tozzi}, {Gobat}, {Santos},
  {Nonino}, {Demarco}, {Lidman}, {Mullis}, {Strazzullo}, {B{\"o}hringer},
  {Fassbender}, {Dawson}, {Tanaka}, {Jee}, {Ford}, {Lamer}, \&
  {Schwope}}]{Rosati2009}
{Rosati}, P., {Tozzi}, P., {Gobat}, R., {et~al.} 2009, \aap, 508, 583

\bibitem[{{Santos} {et~al.}(2015){Santos}, {Altieri}, {Valtchanov}, {Nastasi},
  {B{\"o}hringer}, {Cresci}, {Elbaz}, {Fassbender}, {Rosati}, {Tozzi}, \&
  {Verdugo}}]{Santos2015}
{Santos}, J.~S., {Altieri}, B., {Valtchanov}, I., {et~al.} 2015, \mnras, 447,
  L65

\bibitem[{{Sarazin}(1986)}]{Sarazin1986}
{Sarazin}, C.~L. 1986, Reviews of Modern Physics, 58, 1

\bibitem[{{Sartoris} {et~al.}(2016){Sartoris}, {Biviano}, {Fedeli}, {Bartlett},
  {Borgani}, {Costanzi}, {Giocoli}, {Moscardini}, {Weller}, {Ascaso},
  {Bardelli}, {Maurogordato}, \& {Viana}}]{Sartoris2016}
{Sartoris}, B., {Biviano}, A., {Fedeli}, C., {et~al.} 2016, \mnras
  [\eprint[arXiv]{1505.02165}]

\bibitem[{{Scodeggio} {et~al.}(1995){Scodeggio}, {Solanes}, {Giovanelli}, \&
  {Haynes}}]{Scodeggio1995}
{Scodeggio}, M., {Solanes}, J.~M., {Giovanelli}, R., \& {Haynes}, M.~P. 1995,
  \apj, 444, 41

\bibitem[{{Siegel}(1956)}]{Siegel1956}
{Siegel}, S. 1956, {Nonparametric statistics for the behavioral sciences}
  (McGraw-Hill Kogakusha, Tokyo)

\bibitem[{{Sodr{\'e}} {et~al.}(1989){Sodr{\'e}}, {Capelato}, {Steiner}, \&
  {Mazure}}]{Sodre1989}
{Sodr{\'e}}, Jr., L., {Capelato}, H.~V., {Steiner}, J.~E., \& {Mazure}, A.
  1989, \aj, 97, 1279

\bibitem[{{Tammann}(1972)}]{Tammann1972}
{Tammann}, G.~A. 1972, \aap, 21, 355

\bibitem[{{Tanaka} {et~al.}(2008){Tanaka}, {Finoguenov}, {Kodama}, {Morokuma},
  {Rosati}, {Stanford}, {Eisenhardt}, {Holden}, \& {Mei}}]{Tanaka2008}
{Tanaka}, M., {Finoguenov}, A., {Kodama}, T., {et~al.} 2008, \aap, 489, 571

\bibitem[{{Tran} {et~al.}(2007){Tran}, {Franx}, {Illingworth}, {van Dokkum},
  {Kelson}, {Blakeslee}, \& {Postman}}]{Tran2007}
{Tran}, K.-V.~H., {Franx}, M., {Illingworth}, G.~D., {et~al.} 2007, \apj, 661,
  750

\bibitem[{{Tran} {et~al.}(2010){Tran}, {Papovich}, {Saintonge}, {Brodwin},
  {Dunlop}, {Farrah}, {Finkelstein}, {Finkelstein}, {Lotz}, {McLure},
  {Momcheva}, \& {Willmer}}]{Tran2010}
{Tran}, K.-V.~H., {Papovich}, C., {Saintonge}, A., {et~al.} 2010, \apjl, 719,
  L126

\bibitem[{{Vulcani} {et~al.}(2012){Vulcani}, {Arag{\'o}n-Salamanca},
  {Poggianti}, {Milvang-Jensen}, {von der Linden}, {Fritz}, {Jablonka},
  {Johnson}, \& {Zaritsky}}]{Vulcani2012}
{Vulcani}, B., {Arag{\'o}n-Salamanca}, A., {Poggianti}, B.~M., {et~al.} 2012,
  \aap, 544, A104

\bibitem[{{Westra} {et~al.}(2010){Westra}, {Geller}, {Kurtz}, {Fabricant}, \&
  {Dell'Antonio}}]{Westra2010}
{Westra}, E., {Geller}, M.~J., {Kurtz}, M.~J., {Fabricant}, D.~G., \&
  {Dell'Antonio}, I. 2010, \pasp, 122, 1258

\bibitem[{{White} {et~al.}(2005){White}, {Clowe}, {Simard}, {Rudnick}, {De
  Lucia}, {Arag{\'o}n-Salamanca}, {Bender}, {Best}, {Bremer}, {Charlot},
  {Dalcanton}, {Dantel}, {Desai}, {Fort}, {Halliday}, {Jablonka}, {Kauffmann},
  {Mellier}, {Milvang-Jensen}, {Pell{\'o}}, {Poggianti}, {Poirier},
  {Rottgering}, {Saglia}, {Schneider}, \& {Zaritsky}}]{White2005}
{White}, S.~D.~M., {Clowe}, D.~I., {Simard}, L., {et~al.} 2005, \aap, 444, 365

\bibitem[{{Whitmore} {et~al.}(1993){Whitmore}, {Gilmore}, \&
  {Jones}}]{Whitmore1993}
{Whitmore}, B.~C., {Gilmore}, D.~M., \& {Jones}, C. 1993, \apj, 407, 489

\bibitem[{{Yee} {et~al.}(1996){Yee}, {Ellingson}, \& {Carlberg}}]{Yee1996}
{Yee}, H.~K.~C., {Ellingson}, E., \& {Carlberg}, R.~G. 1996, \apjs, 102, 269

\bibitem[{{Zabludoff} \& {Franx}(1993)}]{Zabludoff1993}
{Zabludoff}, A.~I. \& {Franx}, M. 1993, \aj, 106, 1314

\bibitem[{{Ziparo} {et~al.}(2014){Ziparo}, {Popesso}, {Finoguenov}, {Biviano},
  {Wuyts}, {Wilman}, {Salvato}, {Tanaka}, {Nandra}, {Lutz}, {Elbaz},
  {Dickinson}, {Altieri}, {Aussel}, {Berta}, {Cimatti}, {Fadda}, {Genzel}, {Le
  Floc'h}, {Magnelli}, {Nordon}, {Poglitsch}, {Pozzi}, {Portal}, {Tacconi},
  {Bauer}, {Brandt}, {Cappelluti}, {Cooper}, \& {Mulchaey}}]{Ziparo2014}
{Ziparo}, F., {Popesso}, P., {Finoguenov}, A., {et~al.} 2014, \mnras, 437, 458

\end{thebibliography}

\end{document}